\documentclass[%
 amsmath,amssymb,
 aps,
]{revtex4-1}
\usepackage{graphicx}
\usepackage{dcolumn}
\usepackage{bm}
\usepackage{epstopdf}

\begin{document}

\preprint{APS/123-QED}

\title{Algebraic Solutions for $U^{BF}(5)-O^{BF}(6)$  Quantum Phase Transition in Odd Mass Number Nuclei}

\author{ M. A. Jafarizadeh}
\email{jafarizadeh@tabrizu.ac.ir}
\affiliation{Department of Theoretical Physics and Astrophysics,
University of Tabriz, Tabriz 51664, Iran.}
\affiliation{Research Institute for Fundamental Sciences, Tabriz 51664, Iran}

\author{ M.Ghapanvari}
\affiliation{Department of Nuclear Physics, University of Tabriz, Tabriz 51664, Iran.}

\author{ N.Fouladi}
\affiliation{Department of Nuclear Physics, University of Tabriz, Tabriz 51664, Iran.}

\date{\today}

\begin{abstract}
The spherical to  gamma-unstable nuclei shape- phase
transition in odd-A nuclei is investigated by using the Dual
algebraic structures and the affine $ \widehat{SU(1,1) } $ Lie Algebra within
the framework of the interacting boson - fermion model. The new
algebraic solution for A-odd nuclei is introduced. In this model,
Single $j =  1/2 $ and $ 3/2 $ fermions are coupled with an
even-even boson core. Energy spectra, quadrupole electromagnetic
transitions and an expectation value of the d-boson number operator
are presented. Experimental evidence for the $U^{BF} (5)-O^{BF}
(6)$ transition in odd -A $Ba$ and $Rh$ isotopes is presented.
The low-states energy spectra and $B(E2)$values for these nuclei
have been also calculated and compared with the experimental data.

\end{abstract}

\maketitle
\section{Introduction}
Quantum Phase Transitions $(QPTs)$ are sudden changes in the
structure of a physical system. Nuclear physics has important
contributions to make to their study because nuclei display a
variety of phases in systems ranging from few to many
particles \cite{1}.The signs of $QPTs$ in nuclear physics are
changes in mass and radius of nuclei and quantities such as level
crossing and electromagnetic transition rates when the number of
protons or neutrons is modified. Phase transition happens in both
even-even and $odd-A$ nuclei. Phase transitions investigations
have been mostly performed on even-even systems \cite{2} within
the framework of the interacting boson model $(IBM)$
 \cite{2,3},which describe nuclei in terms of correlated pairs of
nucleons with $L = 0$,$ 2$ treated as bosons $(s,d   \,bosons)$.
The IBM Hamiltonian has exact solutions in three dynamical
symmetry limits ($U(5)$,$ SU(3)$ , and $ O(6)$). These situations
correspond to the spherical, axially deformed, and
gamma-unstable ground state shapes, respectively.
The transition between the $U(5)$-$SU(3)$ limits is a first-order
shape-phase transition while a second-order shape-phase
transition occurs between the $ U(5)$ and $O(6)$ limits
\cite{4,5,6} . During a transition from one limit to another, meet
the points in which potential has flat behavior. These points
are called critical point.Recently Iachello introduced the
so-called critical point symmetries in the framework of the
collective model for even-even nuclei. The critical point from
spherical to $ \gamma -unstable $ shapes, called $E(5)$ \cite{3,6},
the critical point from spherical to axially deformed shapes,
called $ X(5)$ \cite{4,6}, and the critical point from axially
deformed shapes to triaxial shapes, called Y(5) \cite{6,7}. Phase
transitions is also investigate in odd-A nuclei within the
framework of interacting boson - fermion model $(IBFM)$\cite{8},
which describe nuclei in terms of correlated pairs, with $L = 0,
2$ ($s$ , $d $ bosons), and unpaired particles of angular
momentum $j$ $(j fermions)$.Studies of $QPTs$ in odd-even nuclei
were implicitly initiated years ago by Scholten and Blasi
\cite{9}. Several explicit studies have recently been made by
Alonso et al. \cite{10,11,12} and by Boyukata et al. \cite{13},
who also have suggested a simple form of the IBFM Hamiltonian,
particularly well-suited to study $QPTs $ in odd-even nuclei
because of their supersymmetric properties. Similar to even-even
nuclei, also exists critical point for odd-even and even-odd
nuclei but in this case, critical points show with
$E({5}/{\sum_{j}{2j+1}})$ and $X({5}/{\sum_{j}{2j+1}})$ that $j$
is the angular momentum of single nucleon. Iachello \cite{14,15}
has been the case of a $ j (3/2)$ fermion coupled to a boson core
that undergoes a transition from spherical to $\gamma -unstable$
shapes. At the critical point, an elegant analytic solution,
called $ E(5/4)$, has been obtained starting from the Bohr
Hamiltonian \cite{14}.

In this study, we investigate the transition $U^{BF} (5)-O^{BF}
(6) $ in $odd - A $ nuclei. The new algebraic solution for A-odd
nuclei is introduced. For this transition only the boson core
experience the transition and fermion with $j = 1/2$ and $ 3/2 $
coupled to boson core. We evaluate exact solutions for eigenstate
and energy eigenvalues for transitional region in the IBFM by
using the Dual algebraic structure for the two level pairing
model that based on Richardson - Gaudin method and changing the
control parameter that based on affine $ \widehat{SU(1,1) } $ lie algebra. In
order to the investigation of phase transition, we calculate
observables such as level crossing, expectation values of the
d-boson number operator, ground - state energy and its first
derivative  . The low-lying states of $ _{56}^{127-137}Ba$ and $
_{45}^{101-109}Rh$ isotopes have been studied within suggested
model. The results of calculations for these nuclei will present
for energy levels and transitions probabilities, two neutron
separation energies and will compare with the corresponding the
experimental data.
\\This paper is organized as follows: Section 2
briefly summarizes theoretical aspects of transitional
Hamiltonian and affine $ \widehat{SU(1,1) } $ algebraic technique. Sections 3
and 4 include the results and Experimental evidence and sect. 5
is devoted to the summary and some conclusions.

\section{Theoretical framework}
The $ SU(1,1)$ Algebra has been explained in detail in Refs
\cite{16,17,18}.The $ SU(1,1)$ algebra is produced by $ S^{\nu}$,
 $ {\nu=0} $ and ${\pm} $ , which satisfies the following
commutation relations
\begin{equation}
 [S^{0},S^{ \pm }]={\pm}S^{\pm}      \quad     ,    \,\,\,    [S^{+},S^{-} ]=-2S^{0}
\end{equation}
The quadratic Casimir operator of $ SU(1,1)$ can be written as
\begin{equation}
\hat{C } _{2}=S^{0} (S^{0}-1)-S^{+} S^{-}
\end{equation}
The basis states of an irreducible representation $(irrep)$ $
SU(1,1)$ ,$ {|k{\mu}\rangle}$,\, are determined by a single
number $k$ , $k$ where can be any positive number and
${\mu}=k,k+1,...$ Therefore\cite{17,18},
\begin{equation}
\hat{C } _{2}{(SU(1,1))}{|k{\mu}\rangle}=k(k-1){|k{\mu}\rangle}
\quad\quad  ,   \quad\quad
S^{0}{|k{\mu}\rangle}={\mu}{|k{\mu}\rangle}
\end{equation}
In IBM , the generators of $ SU^{d} (1,1)$ generated by the
d-boson pairing algebra
\begin{equation}
S^{+}(d)=\frac{1}{2} (d^{+}.d^{+} ) \quad\quad,S^{-}(d)=\frac{1}{2}
(\widetilde{d}.\widetilde{d} ), \quad\quad S^{0}(d)=\frac{1}{4}{\sum_{\nu}
({d_{\nu}^{+}d_{\nu}+d_{\nu}d_{\nu}^{+}}})=\frac{1}{4}(2\hat{n } _{d}+5)
\end{equation}
Similarly, s- boson pairing algebra forms another $ SU^{s} (1,1)$
algebra generated by
\begin{equation}
 S^{+}(s)=\frac{1}{2} s^{+^2}  \quad , \quad S^{-}(s)=\frac{1}{2} s^{2}
    ,\quad S^{0}(s)=\frac{1}{4} (s^{+} s+ss^{+} )=\frac{1}{4}(2\hat{n } _{s}+1)
\end{equation}
$ SU^{sd} (1,1)$is the s and d boson pairing algebras generated by
\begin{equation}
 S^{+}(sd)=\frac{1}{2} (d^{+}.d^{+}\pm s^{+^2} ) \quad ,
 \quad S^{-}(sd)=\frac{1}{2} (\widetilde{d}.\widetilde{d} \pm s^{2})
 \quad , S^{0}(sd)=\frac{1}{4}{\sum_{\nu}({d_{\nu}^{+}d_{\nu}+d_{\nu}d_{\nu}^{+}})}
  +\frac{1}{4} (s^{+} s+ss^{+} )
  \end{equation}
Because of duality relationships \cite{19,20}, It is known that
the base of $U(5)\supset SO(5) $ and $ SO(6)\supset SO(5) $ are
simultaneously the basis of $ SU^{d} (1,1)\supset U(1) $ and $
SU^{sd} (1,1)\supset U(1) $, respectively. By use of duality
relations \cite{17,19}, the Casimir operators of $SO(5)$ and
$SO(6)$ can also be expressed in terms of the Casimir operators
of $ SU^{d} (1,1)$ and $ SU^{sd} (1,1)$, respectively
\begin{equation}
  \hat{C } _{2}(SU^{d} (1,1))=\frac{5}{16}+\frac{1}{4} \hat{C } _{2}(SO(5) )
\end{equation}
\begin{equation}
  \hat{C } _{2}(SU^{sd} (1,1))=\frac{3}{4}+\frac{1}{4} \hat{C } _{2}(SO(6) )
\end{equation}
The infinite dimensional $ SU(1,1)$ algebra that is generated by
use of \cite{17,18}
\begin{equation}
S_{n}^{\pm}=c_{s}^{2n+1} S^{\pm}(s)+c_{d}^{2n+1}S^{\pm}(d) \quad\quad ,
\quad\quad  S_{n}^{0}=c_{s}^{2n} S^{0} (s)+c_{d}^{2n} S^{0}(d)
\end{equation}
Where $c_{s}$ and $c_{d}$ are real parameters and $ n $ can be
$0,\pm1,\pm2,....$. These generators satisfy the commutation
relations
\begin{equation}
[S_{m}^{0}  ,S_{n}^{\pm} ]=\pm S_{m+n}^{\pm}
\quad\quad     ,     \quad\quad         [S_{m}^{+},S_{n}^{-}
]=-2S_{m+n+1}^{0}
\end{equation}

Then,${S_{m}^{\mu},\mu=0,+,-; m=\pm1,\pm2,...}  $generate an
affine Lie algebra$ \widehat{SU(1,1) } $without central extension.

In odd – A nuclei the Bose $-$ Fermi symmetries are associated with each of the dynamic symmetries of IBM-1\cite{8}. So, the boson algebraic structure will be always taken to be $U^{B}(6)$ , while the fermion algebraic structure will depend on the values of the angular momenta, j, taken into consideration \cite{8}. First we considered the case that a system of N bosons (with L=0, 2) coupled to a fermion with angular momentum j=1/2. The Lattice of algebras in this case  is shown in Fig.1.In Figs 1, 2 , the chain 1 show the state that bosons have $U^{B}(5)$ dynamical symmetric while bosons in chain 2 have $O^{B}(6)$ dynamical symmetric.By
employing the generators of Algebra $ \widehat{SU(1,1) } $ and Casimir operators of subalgebras , the following
Hamiltonian for transitional region between $U^{BF}
(5)-O^{BF}(6)$ limits is prepared

\begin{figure}[htb]
\begin{center}
\includegraphics[height=8cm]{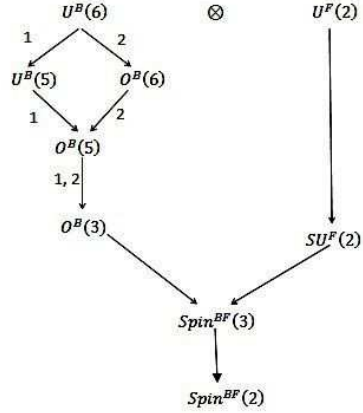}
\caption{The Lattice of algebras in the case that a system of N
bosons (with $L=0, 2$) coupled to a fermion with angular momentum
$j=1/2$.\label{fig:12}}
\end{center}
\end{figure}

\begin{figure}[htb]
\begin{center}
\includegraphics[height=8cm]{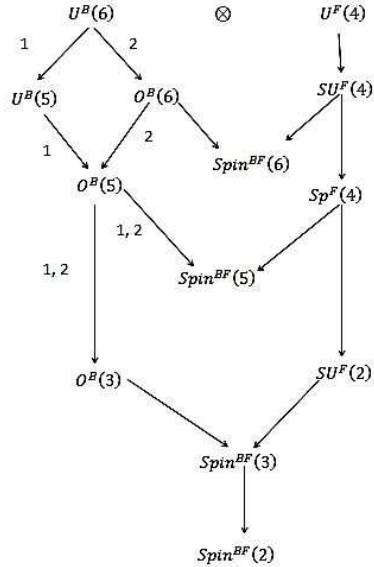}
\caption{ The Lattice of algebras in the case that a system of N
bosons (with $L=0, 2$) coupled to a fermion with angular momentum
$j=3/2$.\label{fig:13}}
\end{center}
\end{figure}

\begin{equation}
\hat{H }=gS_{0}^{+} S_{0}^{-}+\alpha S_{1}^{0}+\beta\hat{C }
_{2}(SO^{B}(5) )+\delta\hat{C } _{2}(SO^{B} (3) )+\gamma\hat{C }
_{2}(spin^{BF} (3) )
\end{equation}
Following this, we considered the state that the odd nucleon being in a j=3/2 shell. The Lattice of algebras in this case  is also shown in Fig.2 .The following
Hamiltonian for state that odd nucleon being in a j=3/2 shell for transitional region  between $U^{BF}
(5)-O^{BF}(6)$ limits   is prepared

\begin{equation}
\hat{H }=gS_{0}^{+} S_{0}^{-}+\alpha S_{1}^{0}+\beta\hat{C }
_{2}(spin^{BF} (5) )+\gamma\hat{C } _{2}(spin^{BF} (3))
\end{equation}
Eqs.(11) and (12) are the suggested Hamiltonians for boson -
fermion systems with $j=1/2  , 3/2$, respectively and $\alpha$
, $\beta$ ,$ \delta$ ,$\gamma$ are real parameters. By considering Eqs.(2),(7),(8), It can be shown that
Hamiltonians (11) and (12) are equivalent with $ O^{BF} (6)$
Hamiltonian when $c_{s}=c_{d}$  and with $ U^{BF} (5)$
Hamiltonian if $c_{s}=0 $ and $c_{d}\neq0 $. Thus, As mentioned because that only the boson core experience the transition and fermion coupled to boson core , the $c_{s}\neq
c_{d}\neq 0$  situation just corresponds to $U^{BF}
(5)\leftrightarrow O^{BF} (6)$ transitional region. In our
calculation, we take  $c_{d}(=1)$ constant value and $c_{s}$
change between 0 and $c_{d}$. For evaluating the eigenvalues of
Hamiltonians (11) and (12) the eigenstates are considered as
\cite{17,18}
\begin{equation}
|k;\nu_{s}\nu n_\Delta LM \rangle={\sum_{n_{i}\in Z}{a_{n_{1}n_{2}....
n_{k}}x_{1}^{n_{1}}x_{2}^{n_{2}}x_{3}^{n_{3}}...x_{k}^{n_{k}}S_{n_{1}}^{+}
S_{n_{2}}^{+}S_{n_{3}}^{+}...S_{n_{k}}^{+.}|lw\rangle}}
\end{equation}
Eigenstates of Hamiltonians (11) and (12) can obtain with using
the Fourier-Laurent expansion of eigenstates and $SU(1,1)$
generators in terms of c-unknown  number parameters $ x_{i}$ with
$ i=1,2,...,k$. It means, one can consider the eigenstates as
\cite{17,18}
\begin{equation}
|k;\nu_{s}\nu n_\Delta LM \rangle= \textit{N}
S_{x_{1}}^{+}S_{x_{2}}^{+}S_{x_{3}}^{+}...S_{x_{k}}^{+}|lw\rangle^{BF}
\end{equation}
Where $ N $ is the normalization factor and
\begin{equation}
S _{x_{i} }^+=\frac {c_{s}}{1-c_{s}^{2} x_{i} } S^{+} (s)+\frac
{c_{d}}{1-c_{d}^{2} x_{i} } S^{+} (d)
\end{equation}
The c-numbers $x_{i}$ are determined through the following set of
equations:
\begin{equation}
\frac {\alpha}{x_{i}}=\frac{g c_{s}^{2} (\nu_{s}+\frac
{1}{2})}{1-c_{s}^{2} x_{i}}+\frac{g c_{d}^{2} (\nu_{d}+\frac
{5}{2})}{1-c_{d}^{2} x_{i}}-{\sum_{j\neq i}{\frac
{2g}{x_{i}-x_{j}}}}
\end{equation}
With $Clebsch- Gordan (CG)$ coefficient, we can calculate lowest
weight state, $|lw\rangle^{BF}$, in terms of boson and fermion
part.For the  j=1/2 case we have:
\begin{equation}
|lw\rangle_{m\pm \frac{1}{2}}^{B}=|N,k_{d}=\frac{1}{2}
(\nu_{d}+\frac{5}{2}),\mu_{d}=\frac{1}{2}
(n_{d}+\frac{5}{2}),k_{s}=\frac{1}{2}
(\nu_{s}+\frac{1}{2}),\mu_{s}=\frac{1}{2} (n_{s}+\frac{1}{2}),L,m
\pm \frac{1}{2}\rangle
\end{equation}
\begin{equation}
|lw\rangle^{BF}=\pm \sqrt{{\frac {L\pm
m+\frac{1}{2}}{(2L+1)}}}|lw\rangle_{m-\frac{1}{2}}^{B}\chi_{+} +
\sqrt{{\frac {L\mp
m+\frac{1}{2}}{(2L+1)}}}|lw\rangle_{m+\frac{1}{2}}^{B}\chi_{-}
\end{equation}
The lowest weight state for the j=3/2 case is calculated as:
\begin{equation}
|lw\rangle_{m\pm \frac{3}{2}}^{B}=|N,k_{d}=\frac{1}{2}
(\nu_{d}+\frac{5}{2}),\mu_{d}=\frac{1}{2}
(n_{d}+\frac{5}{2}),k_{s}=\frac{1}{2}
(\nu_{s}+\frac{1}{2}),\mu_{s}=\frac{1}{2} (n_{s}+\frac{1}{2}),L,m
\pm \frac{3}{2}\rangle
\end{equation}
\begin{eqnarray}
|lw\rangle^{BF}=&C_{m,m-\frac{3}{2},\frac{3}{2}}^{J,L,\frac{3}{2}}|lw\rangle_{m- \frac{3}{2}}^{B}|j=\frac{3}{2},m_{j}=\frac{3}{2}\rangle+C_{m,m-\frac{1}{2},\frac{1}{2}}^{J,L,\frac{3}{2}}|lw\rangle_{m- \frac{1}{2}}^{B}|j=\frac{3}{2},m_{j}=\frac{1}{2}\rangle
\nonumber \\
&+C_{m,m+\frac{1}{2},-\frac{1}{2}}^{J,L,\frac{3}{2}}|lw\rangle_{m+ \frac{1}{2}}^{B}|j=\frac{3}{2},m_{j}=-\frac{1}{2}\rangle
+C_{m,m+\frac{3}{2},-\frac{3}{2}}^{J,L,\frac{3}{2}}|lw\rangle_{m+ \frac{3}{2}}^{B}|j=\frac{3}{2},m_{j}=-\frac{3}{2}\rangle
\end{eqnarray}
The $C_{m,m_{L},m_{j}}^{J,L,j} $symbols represent Clebsch-Gordan coefficients.
where
\begin{equation}
S_{n}^{0} |lw\rangle^{BF} =\Lambda_{n}^{0} |lw\rangle^{BF}
\end{equation}
\begin{equation}
\Lambda_{n}^{0} = c_{s}^{2n}(\nu_{s}+\frac {1}{2})\frac
{1}{2}+c_{d}^{2n}(\nu_{d}+\frac {5}{2})\frac {1}{2}
\end{equation}
The eigenvalues of Hamiltonians (11), (12) can then be expressed;
\begin{equation}
E^{(k) }=h^{(k) }+\alpha \Lambda_{1}^{0}+\beta
\nu_{d}(\nu_{d}+3)+\delta L(L+1)+\gamma J(J+1)
\end{equation}
\begin{equation}
E^{(k) }=h^{(k) }+\alpha \Lambda_{1}^{0}+\beta(\nu_{1}
(\nu_{1}+3)+\nu_{2} (\nu_{2}+1) )+\gamma J(J+1))
\end{equation}
\begin{equation}
h^{(k) }=\sum_{i=1}^{k}{\frac {\alpha}{x_{i}}}
\end{equation}
The quantum number $(k)$ is related to the total boson number $N$
by
$$ N=2k+\nu_{s}+\nu_{d}$$
In order to obtain the numerical results for energy spectra
$(E^{(k) } )$ of considered nuclei, a set of non-linear $
Bethe-Ansatz$ equations (BAE) with k- unknowns for k-pair
excitations must be solved \cite{17,18} also constants of
Hamiltonian with least square fitting processes to experimental
data is obtained. To this aim, we have changed variables as
$$ C=\frac  {c_{s}}{c_{d}} \leq 1    ,    g=1       ,    y_{i}=c_{d}^{2} x_{i} $$
So, the new form of Eq.(16) would be
\begin{equation}
\frac {\alpha}{y_{i}}=\frac{ C^{2} (\nu_{s}+\frac
{1}{2})}{1-C^{2} y_{i}}+\frac{ (\nu_{d}+\frac
{5}{2})}{1- y_{i}}-{\sum_{j\neq i}{\frac {2}{y_{i}-y_{j}}}}
\end{equation}
To calculate the roots of $Bethe-Ansatz$ equations (BAE) with
specified values of $ \nu_{s} $ and $\nu_{d} $, we have solved
Eq. (24) with definite values of C and $ \alpha $
\cite{16}.Then, we carry out this procedure with different values
of C and $ \alpha $  to give energy spectra with minimum
variation in compare to experimental values \cite{21};
$$ \sigma = (\frac {1}{N_{tot}}\sum _{i,tot}{|E_{exp}(i)-E_{Cal}(i)|^2})^{\frac {1}{2}}$$
($N_{tot}$ the number of energy levels where included in the
fitting processes). The method for optimizing the set of
parameters in the Hamiltonian $(\beta,\gamma,\delta)$  includes
carrying out a least-square fit (LSF) of the excitation energies
of selected states \cite{16}
\section{Results:}
This section presented the calculated phase transition
observables such as level crossing, ground - state energy and the
derivative of the energy, expectation values of the d-boson
number operator and energy differences.
\subsection{energy spectrum and level crossing}
To display how the energy levels change as a function of the
control parameter C and the total number of bosons N, the lowest
energy levels as a function of C for $N=10, 20$  bosons are shown
in Fig.3 and Fig.4, where in Fig.3 other fixed parameters are $
\alpha=1000 $, $\beta=-57 $, $ \delta=41 $, $\gamma=36 $ and Fig.4 obtained
with $ \alpha=1000$, $ \beta=6.5 $, $\gamma=-22$. Figs show how the
energy levels as a function of the control parameter C evolve
from one dynamical symmetry limit to the other . It can be seen
from Figs that numerous level crossings occur; especially in the
region around $C \geq 0.7$. The crossings are due to the fact
that $ \nu_{d} $, $O(5)$ quantum number called seniority, is
preserved along the whole path between $ O(6)$ and $U(5)$
\cite{22}. With increasing N,  level crossing increase that in
Fig.4 clearly shows.

\begin{figure}[htb]
\begin{center}
\includegraphics[height=6cm]{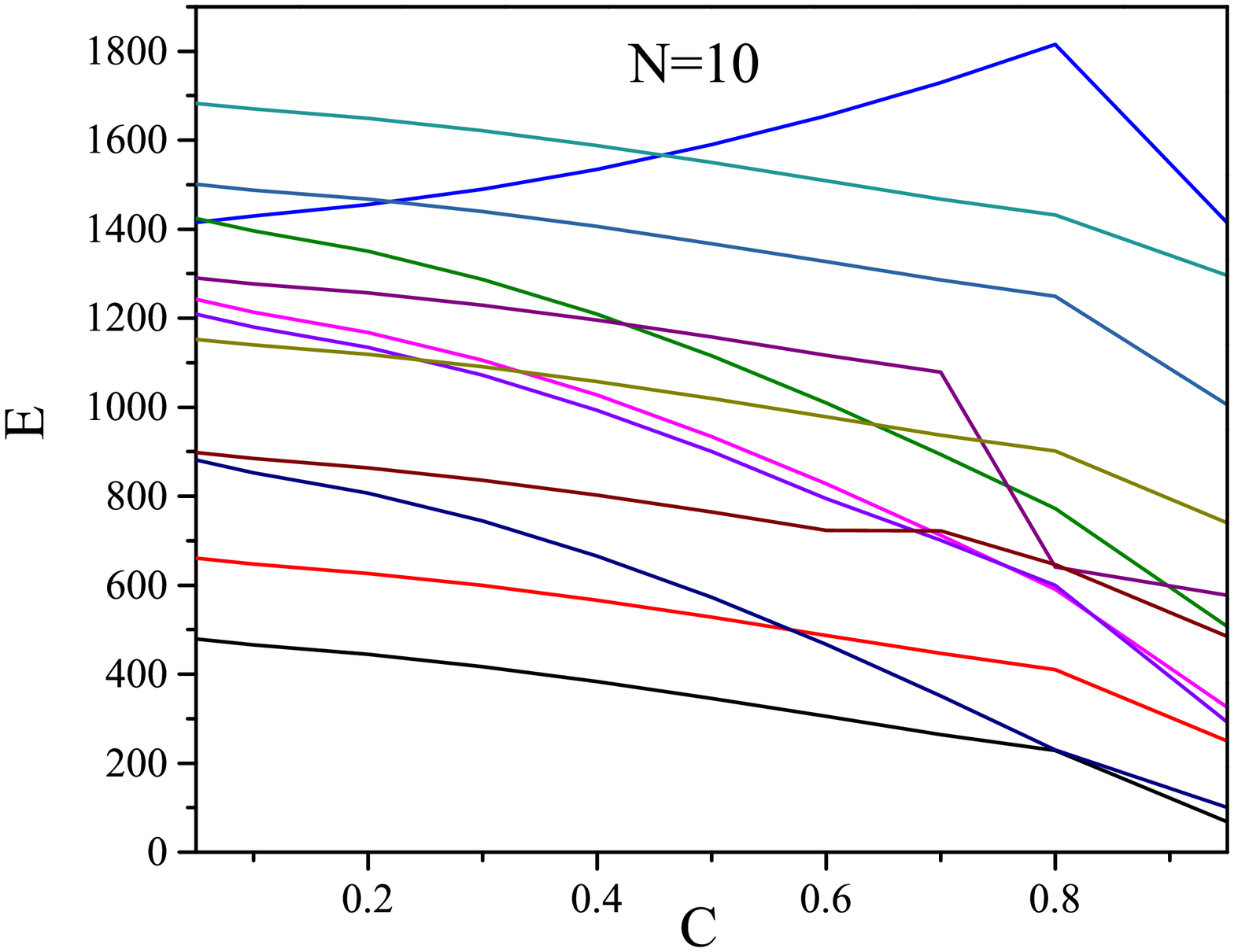}
\includegraphics[height=6cm]{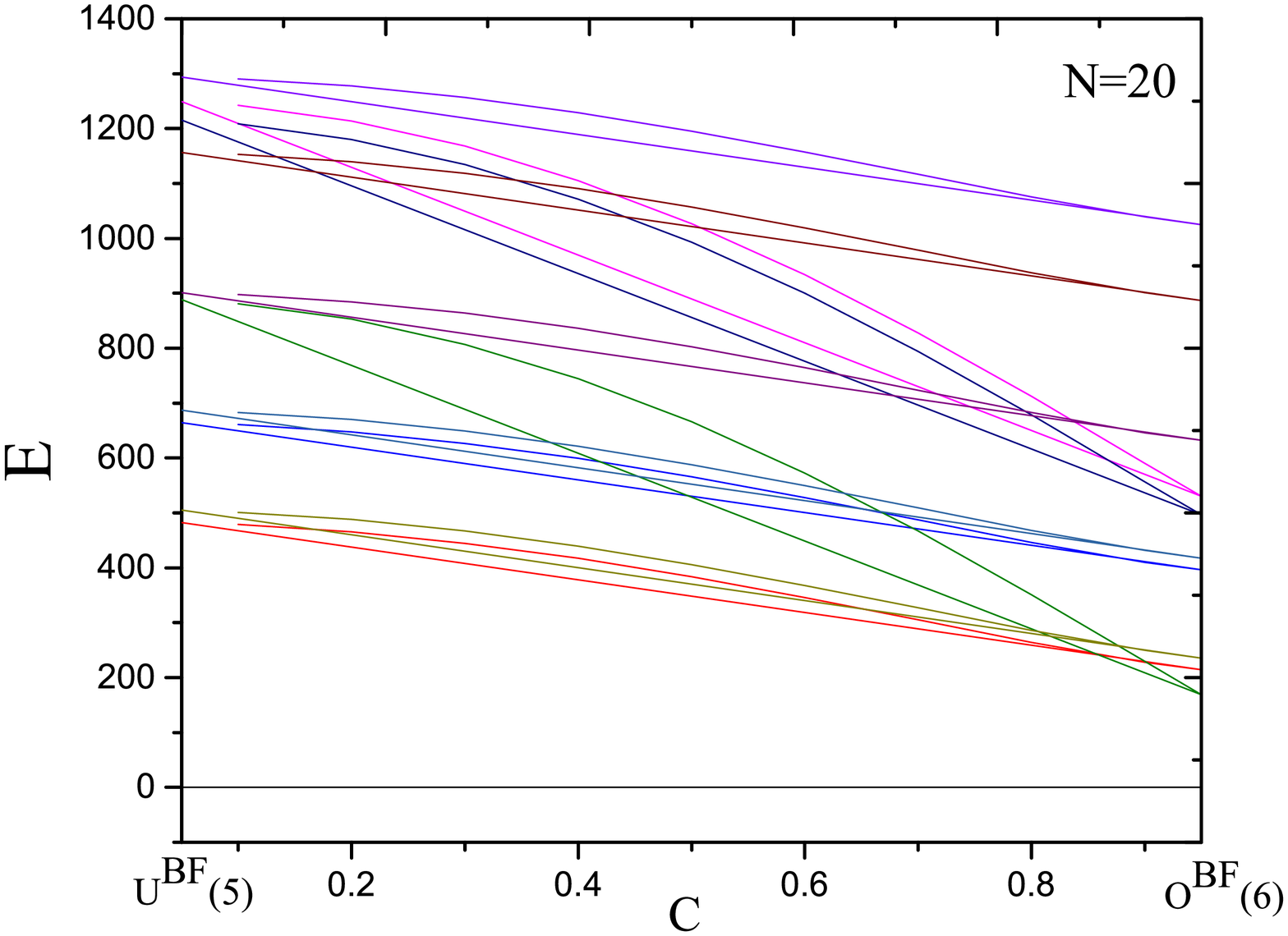}
\caption{Energy levels as a function of the control parameter $C$
in the Hamiltonian (11) for $N=10, 20$ bosons
with$\alpha=1000,\beta=-57,\delta=41,\gamma=36$.\label{fig:1}}
\end{center}
\end{figure}
\begin{figure}[htb]
\begin{center}
\includegraphics[height=6cm]{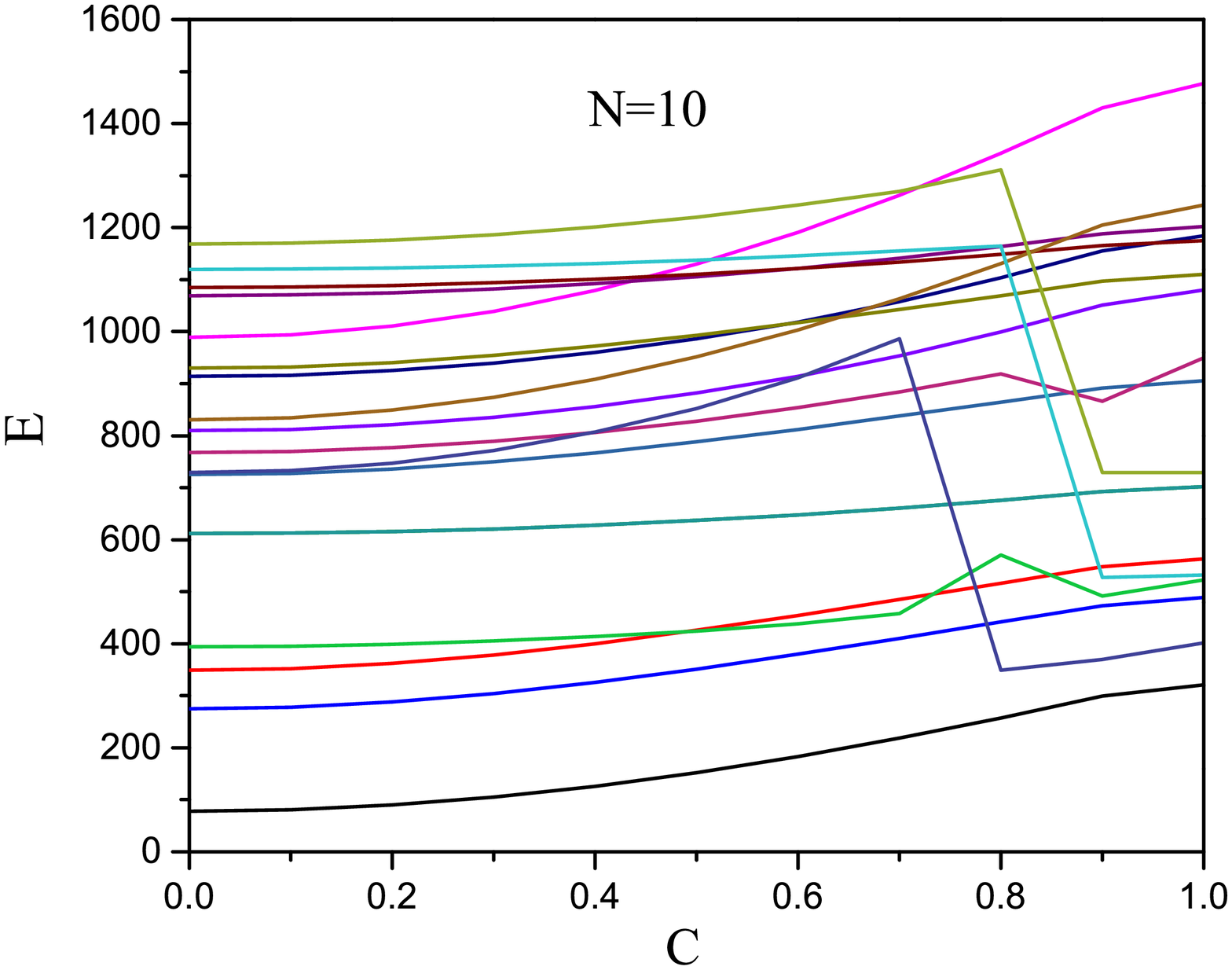}
\includegraphics[height=6cm]{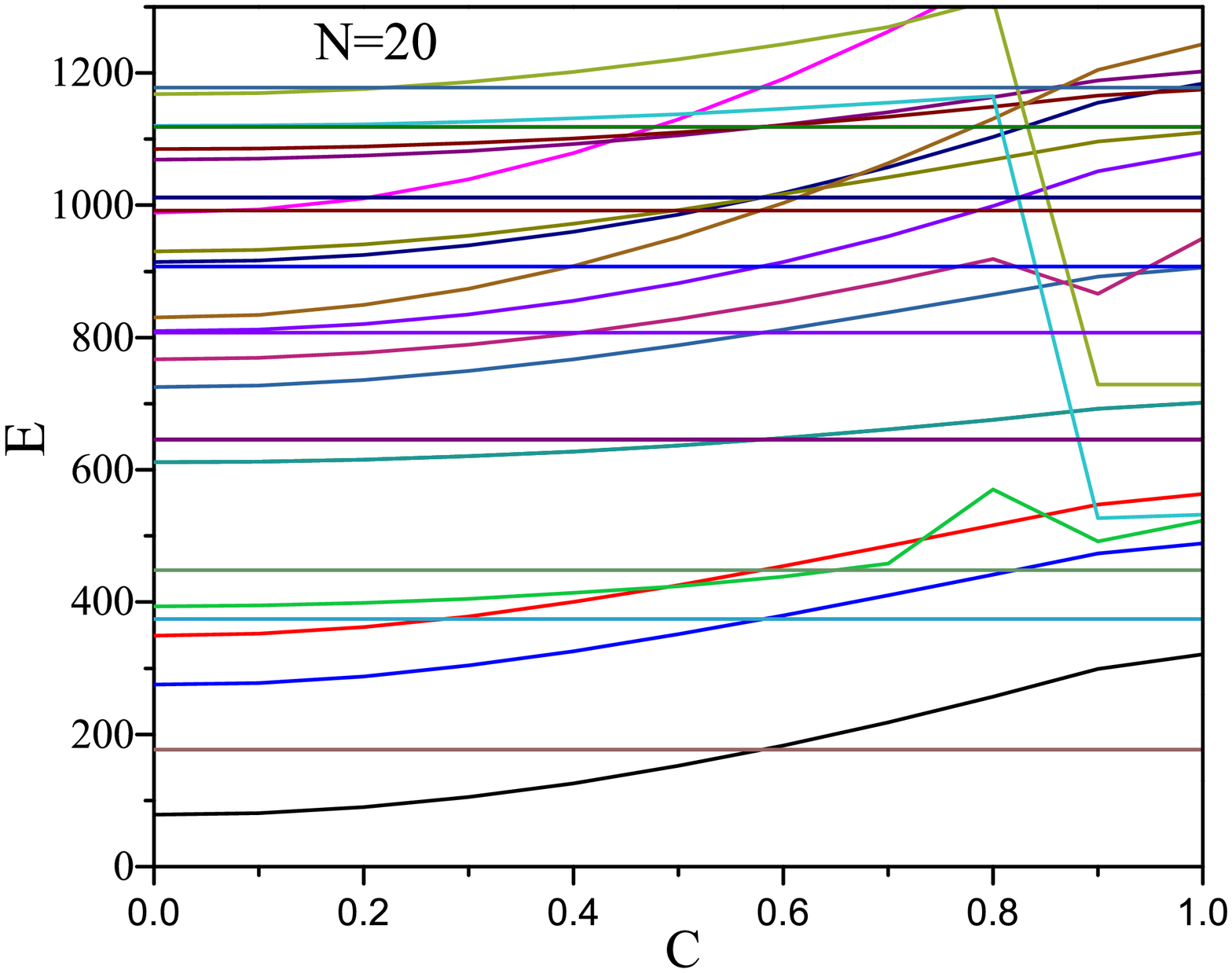}
\caption{Same as Fig.3 but with Hamiltonian (12)
with$\alpha=1000,\beta=6.5,\gamma=-22$.\label{fig:2}}
\end{center}
\end{figure}
\begin{figure}[htb]
\begin{center}
\includegraphics[height=6cm]{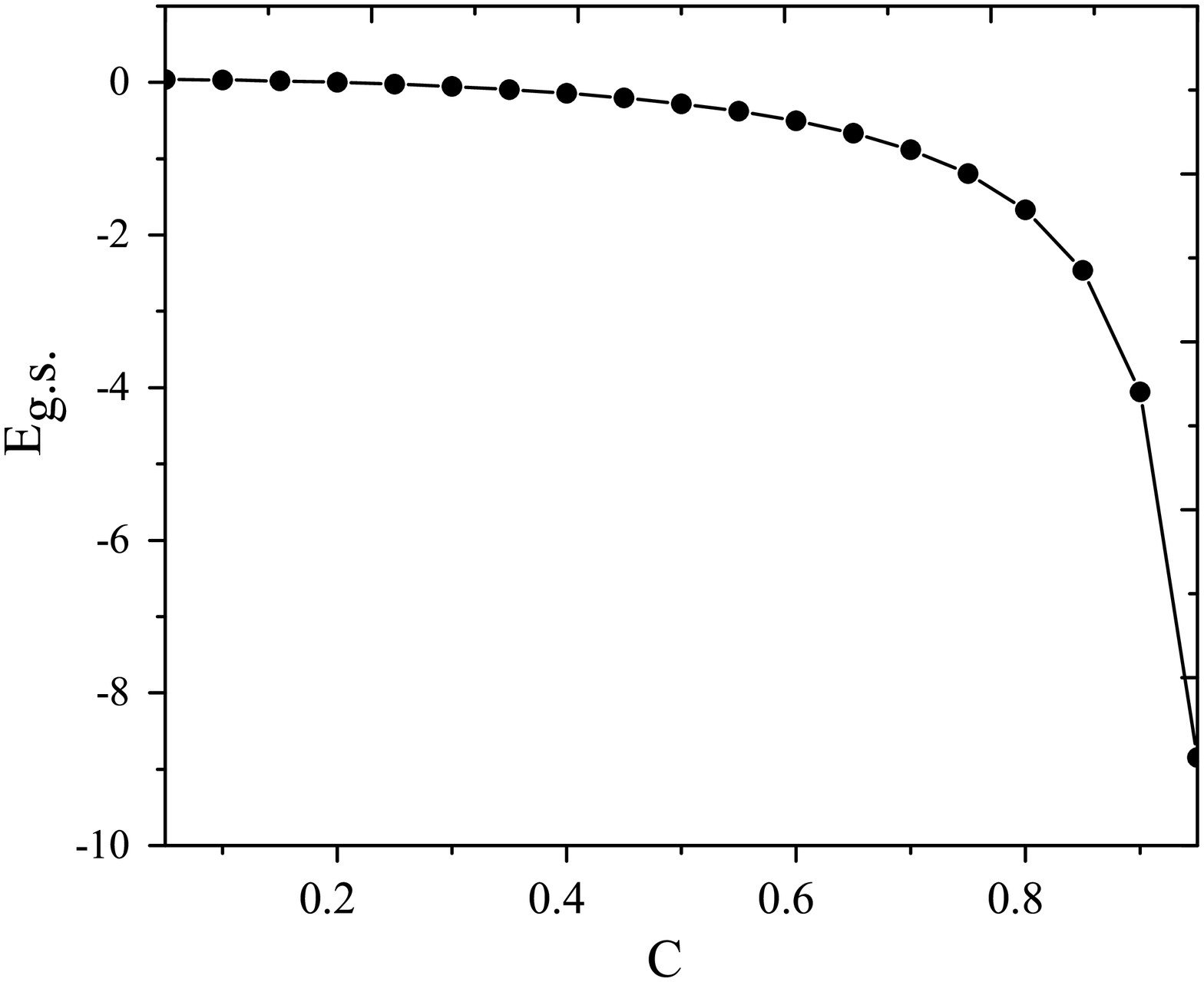}
\includegraphics[height=6cm]{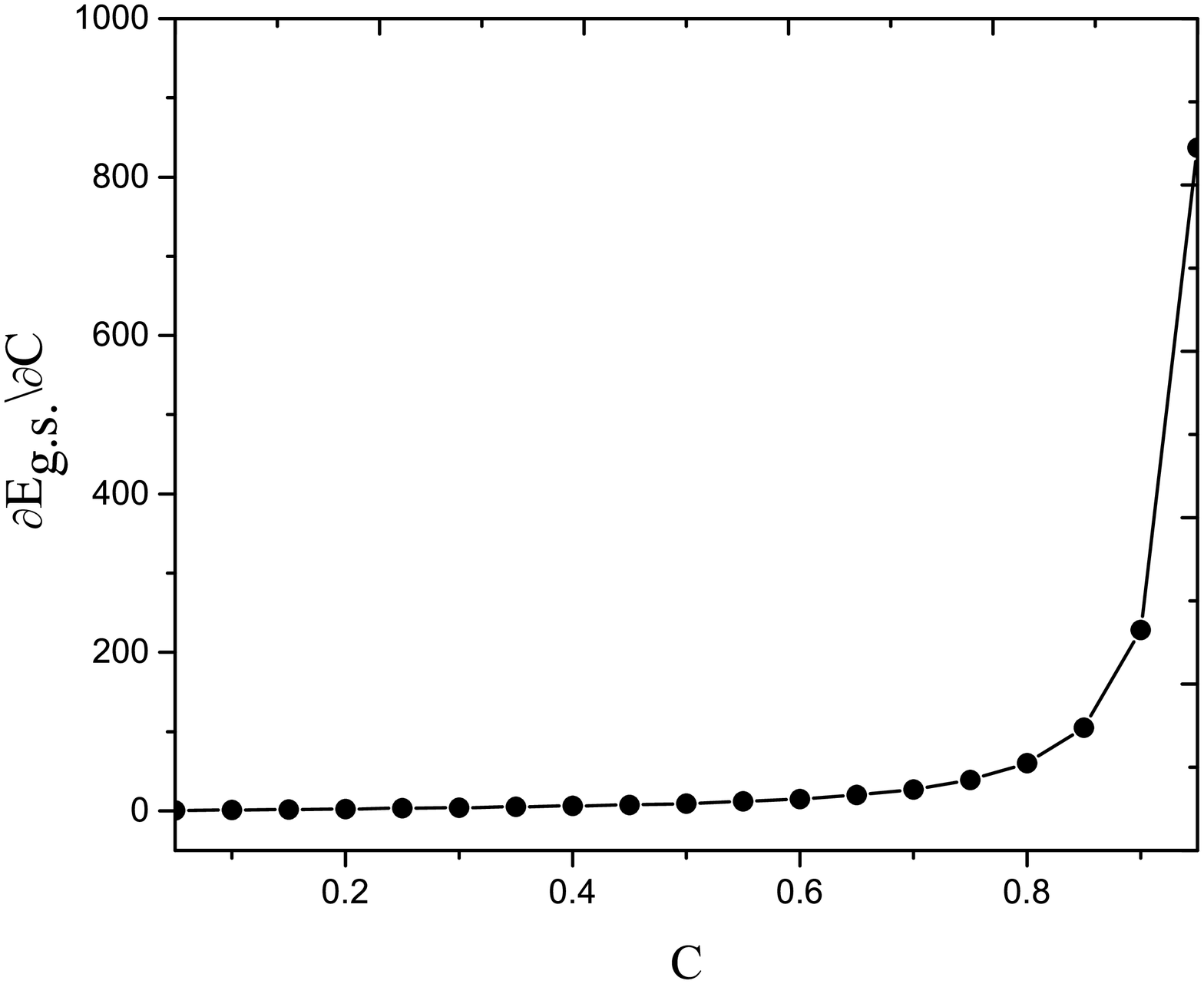}
\caption{The ground-state energy (left panel) and derivative of
the ground-state energy (right panel) are presented as a function
of the control parameter $C$ for a system with $10$
bosons.\label{fig:3}}
\end{center}
\end{figure}

\subsection{Ground state energy}
The ground state energy is an important observable of phase
transition. So, we calculated the ground - state energy
,$E_{g.s.}$ and its first  derivative ,$\frac {\partial
E_{g.s.}}{\partial C}$. Fig.5 shows  changing of the ground -
state energy  and its first  derivative  versus the control
parameter C. Both operators, $E_{g.s.}$ and $\frac {\partial
E_{g.s.}}{\partial C}$, are approximately zero in one phase and
different from zero in the other phase. Since low N bosons is
chosen, it is not possible to distinguish whether the transition
is first or second order such as done in the even-even case
\cite{23,24}.
\subsection{expectation values of the d-boson number operator}
An appropriate quantal order parameter is:
$$ \langle \hat{n_{d}}\rangle=\frac {\langle \psi |\hat{n_{d}}|\psi\rangle}{N}$$
In order to obtain $\langle \hat{n_{d}}\rangle$, we act
$s_{m}^{0}$ on the eigenstate, $|k;\nu_{s}\nu n_\Delta LM \rangle$
\begin{equation}
\langle \hat{n_{d}}\rangle=\frac {2\Lambda_{1}^{0}-2C^{2}
\Lambda_{0}^{0}+2k(y_{1}^{-1}-C)} {1-C^{2}}-\frac {5}{2N}
\end{equation}

\begin{figure}[htb]
\begin{center}
\includegraphics[height=6cm]{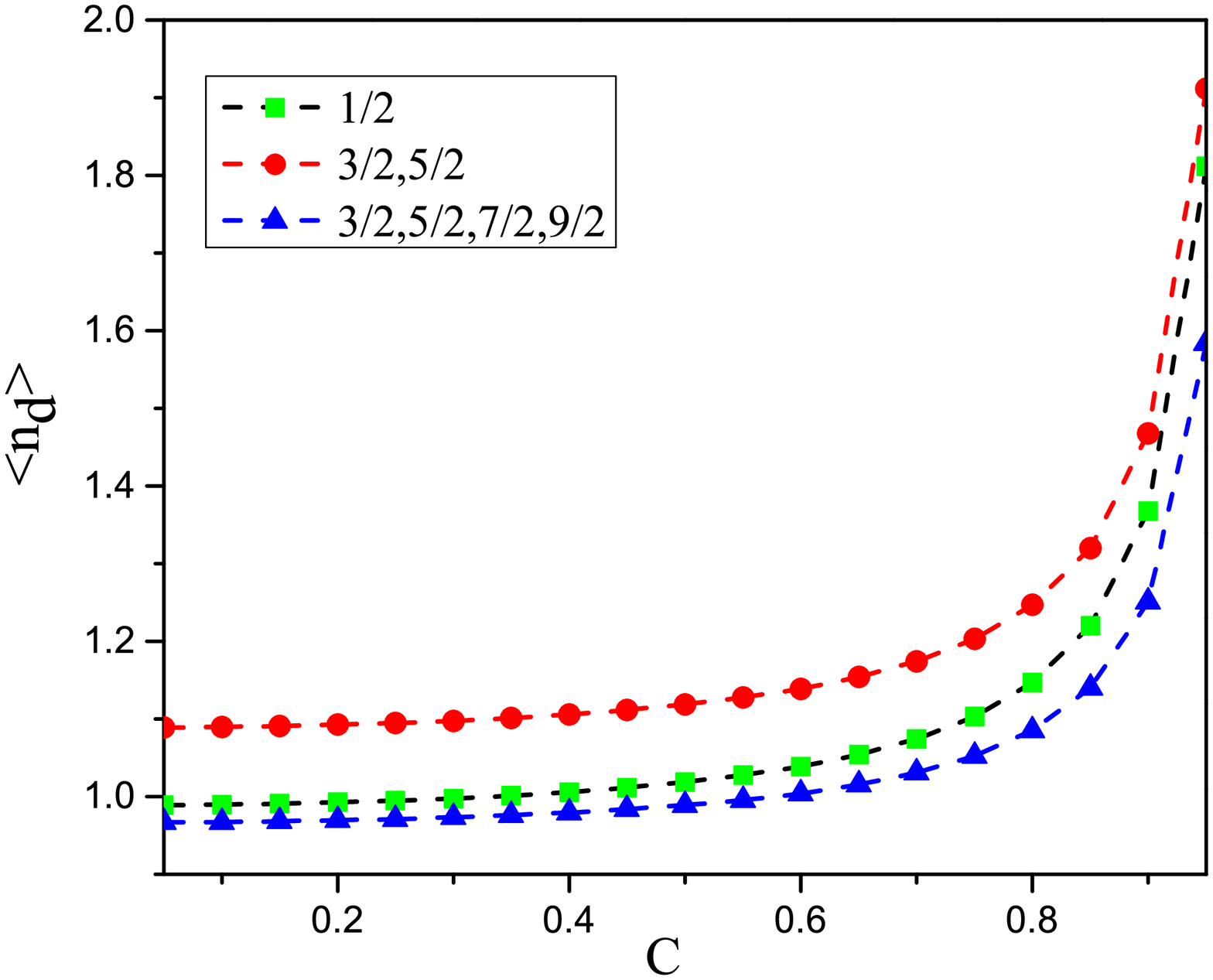}
\includegraphics[height=6cm]{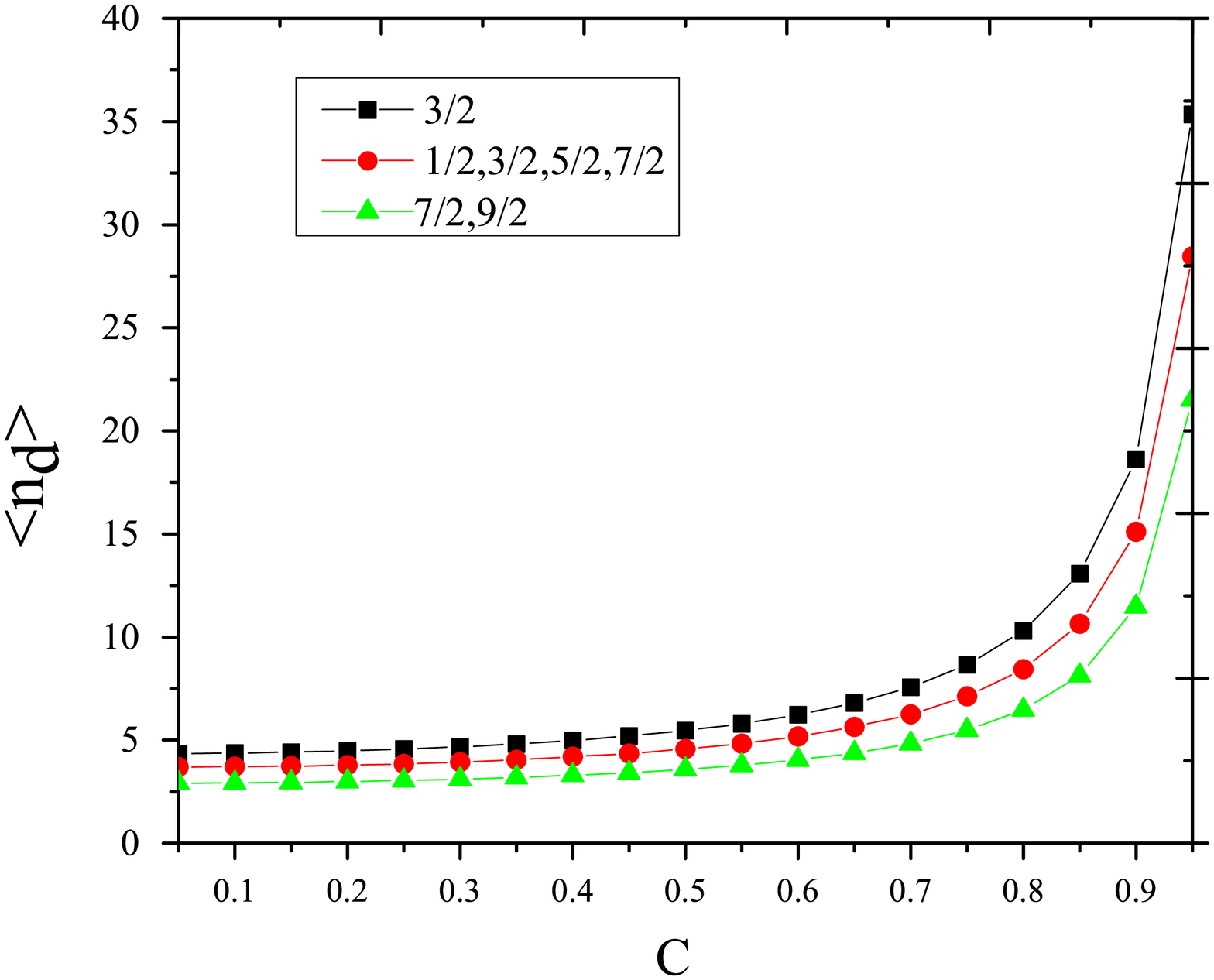}
\caption{the expectation values of the $d$-boson number operator
for the lowest states as a function of $C$ control parameter for a
$j=1/2$ particle coupled to a system of $(s, d)$ bosons undergoing
a $U(5)-O(6)$ transition (left panel) and single fermion with
$j=3/2$ coupled to a system of $(s, d)$ bosons (right
panel).\label{fig:4}}
\end{center}
\end{figure}

Fig.6 shows the expectation values of the d-boson number operator
for the lowest states as a function of C control parameter for
N=10 bosons. Fig.6 displays that the expectation values of the
number of d bosons for each J, $n_{d}$, remain approximately
constant for $ C <0.45$ and only begin to change rapidly for $ C
>0.45$. The near constancy of $n_{d}$ for $C <0.45$, is a obvious
indication that $ U(5)$ dynamical symmetry preserves in this
region to a high degree and also the $n_{d}$ values change
rapidly with C over the range $0.65\leq C \leq 1$. It can be seen
from Fig.6 that in due to the presence of the fermion, the
transition is made sharper for states that a $j=1/2$ particle
coupled to a system of $(s, d)$ bosons undergoing a $U(5)-O(6)$
transition (left panel) while is made smoother for a $j=3/2$
particle coupled to a system of $(s, d)$ bosons (right panel).

\subsection{energy differences}
Fig.7 displays continues energy differences in terms of control
parameter, C, for states that a $j=1/2$ and $j=3/2$ particle
coupled to a system of (s, d) bosons. Fig.7 shows that during
transition from one limit to another exist the points that energy
is minimum or maximum near the critical point.

\begin{figure}[htb]
\begin{center}
\includegraphics[height=6cm]{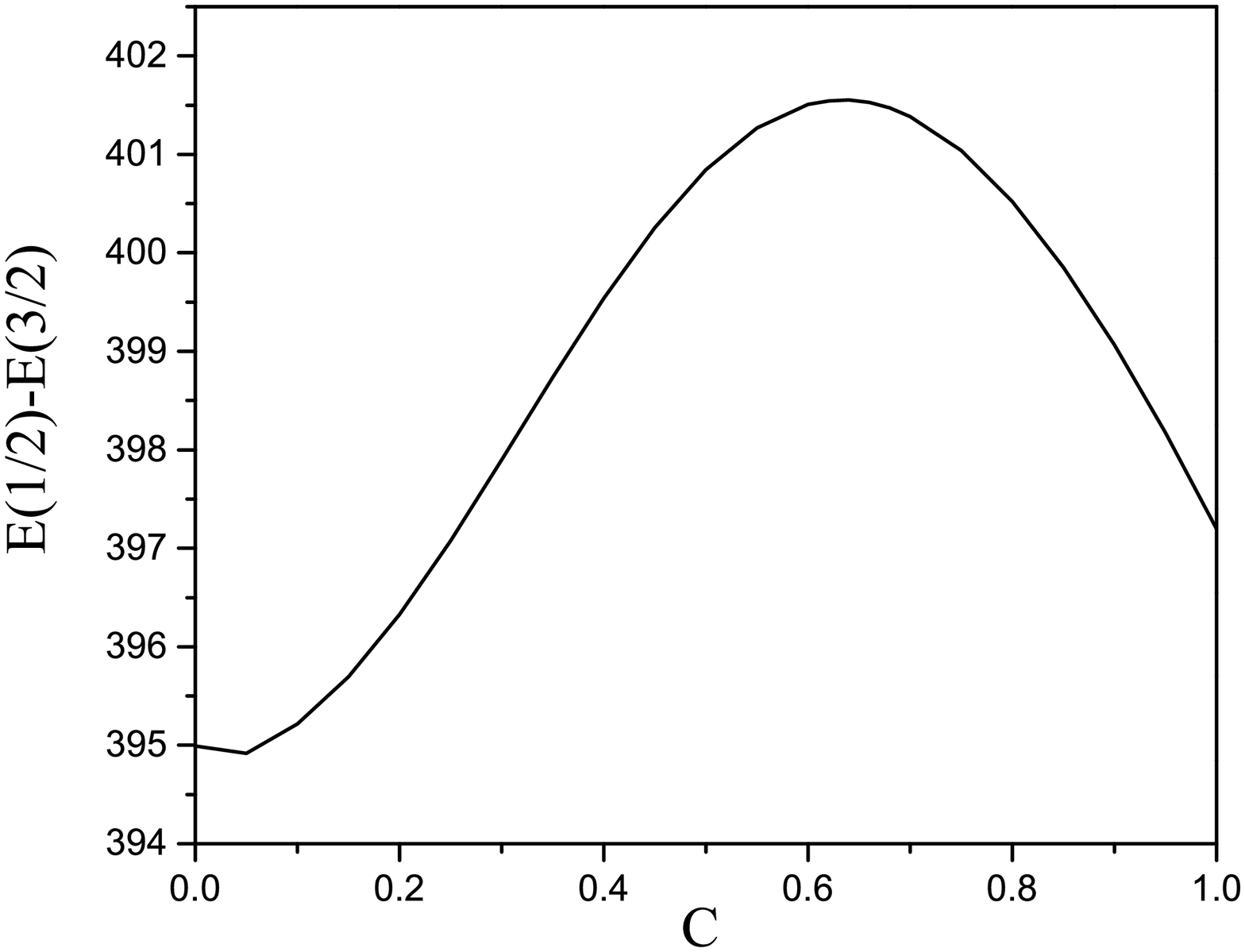}
\includegraphics[height=6cm]{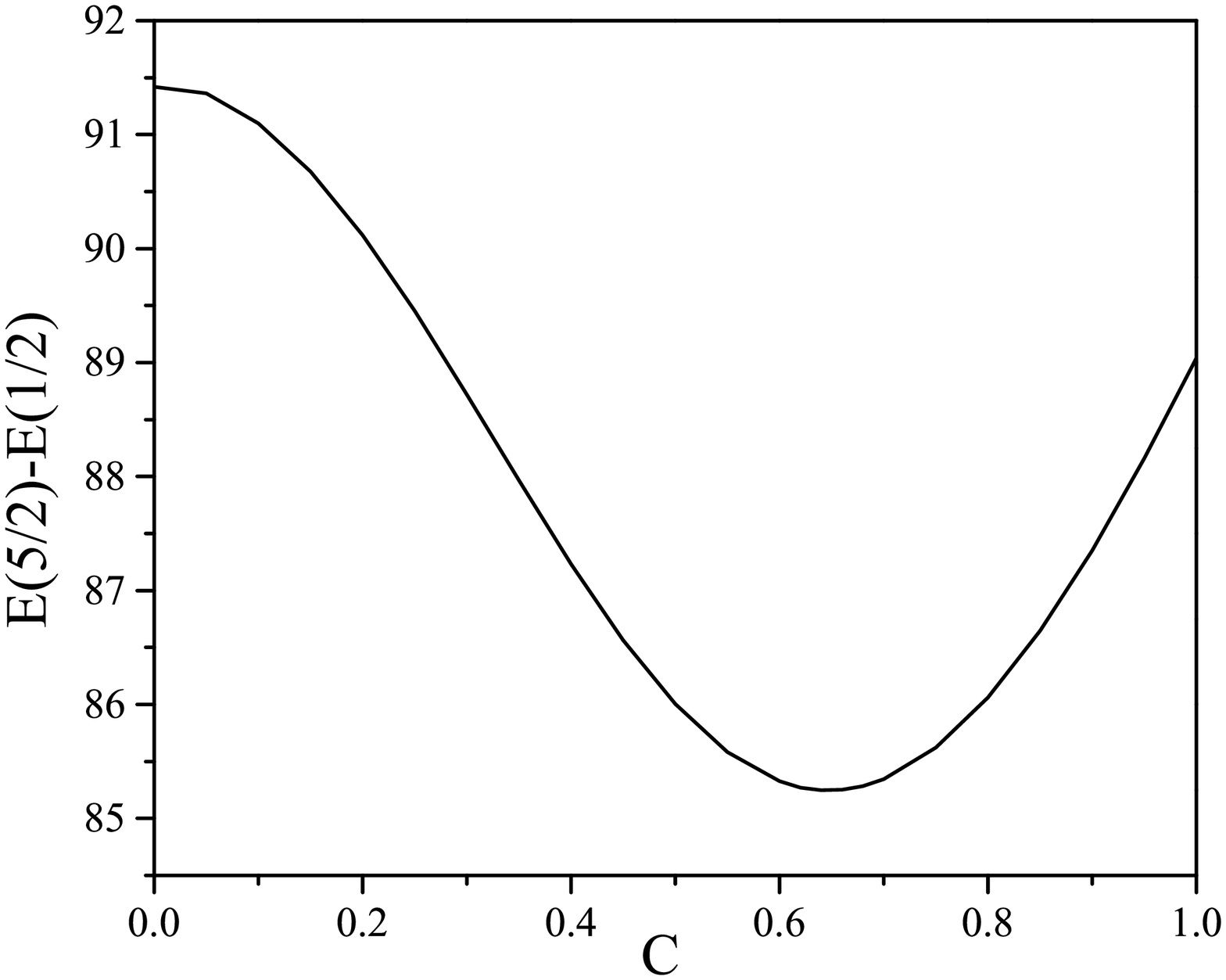}
\includegraphics[height=6cm]{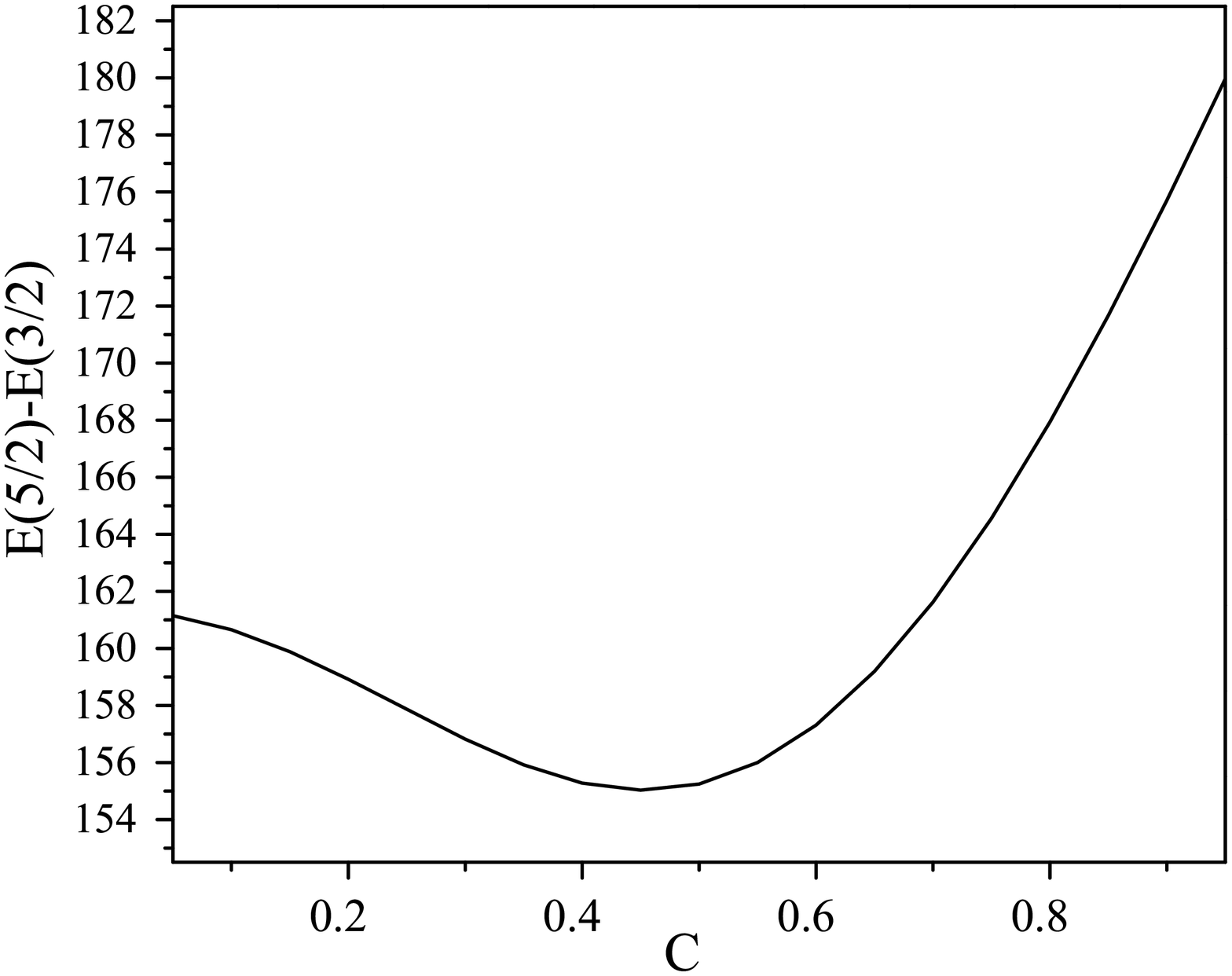}
\includegraphics[height=6cm]{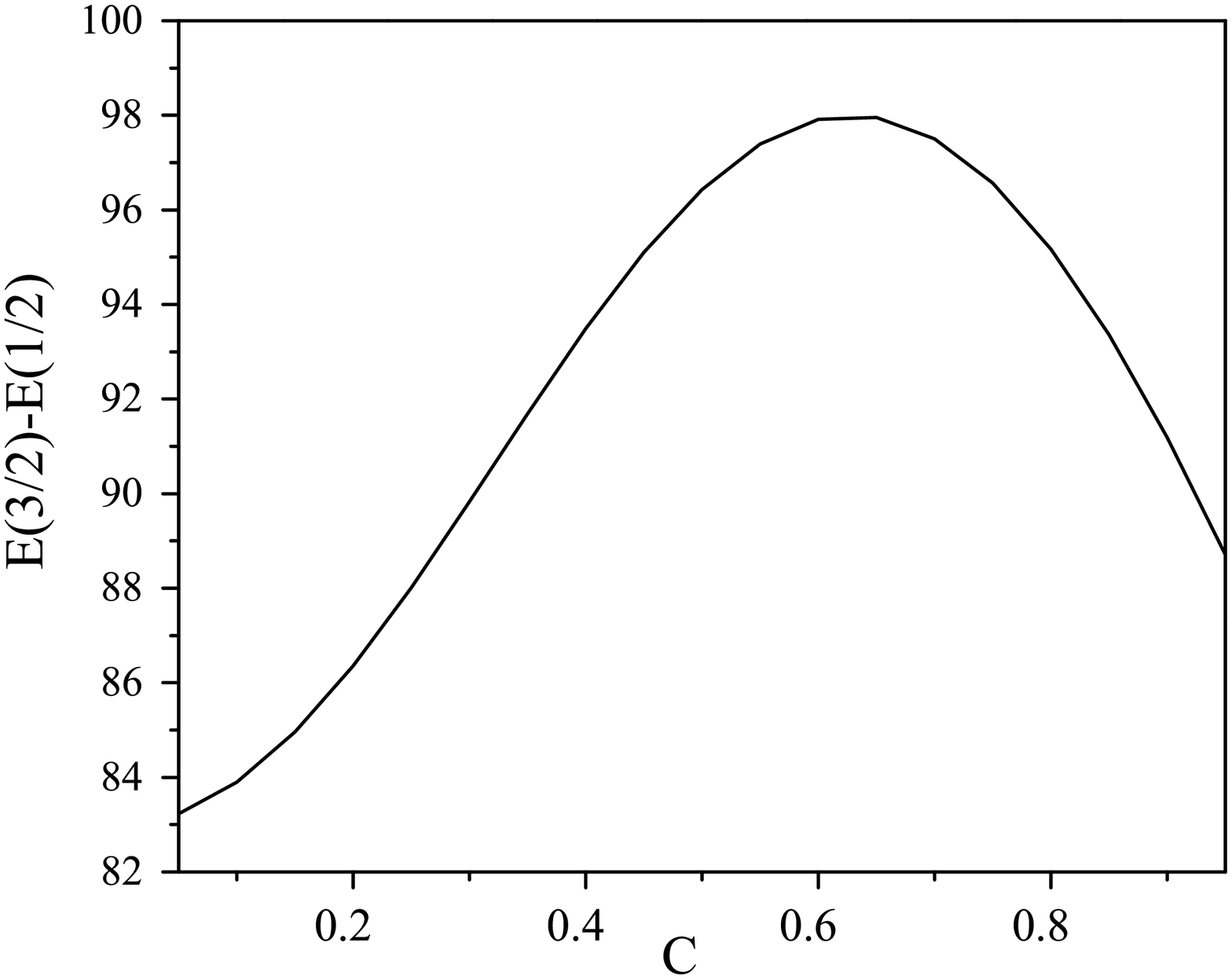}
\caption{Continues energy differences in terms of control
parameter, $C$, for states that a $j=3/2$ (top) and
$j=1/2$(bottom) particle coupled to a system of $(s, d)$
bosons.\label{fig:5}}
\end{center}
\end{figure}

\section{Experimental evidence}
This section presented the calculated results of low-lying states
of the odd -A Ba and Rh isotopes. The results include energy
levels and the B( E2)values and two-neutron separation energies.
\subsection{energy spectrum}
Nuclei in the mass regions around $A  \sim  100$ \cite{25,26} and
$130$ \cite{27} have transitional characteristics intermediate
between spherical and gamma-unstable shapes . The
theoretical and experimental studies of energy spectra done in
refs.\cite{26,28,29} show Rh and Ba isotopes have $
U(5)\leftrightarrow O(6) $  transitional characteristics. .The
possible occurrence of this symmetry in $_{56}^{135}Ba$ has been
recently suggested \cite{28}.The negative parity states in the odd
- even nuclei Rh are built mainly on the $ 2p_{\frac {1}{2}} $
shell model orbit\cite{26}. The single - particle orbits $ 1g
_{\frac {7}{2}}, 2d_{\frac {5}{2}}, 2d_{\frac {3}{2}}   and
3s_{\frac {1}{2}}  $  establish the positive parity states in
odd-mass Ba isotopes\cite{29}. In this study, a simplifying
assumption is made that single particle states are built on the $
j=1/2 $ and $ j=3/2 $. We therefore analyze the negative parity
states of the odd-proton nuclei, $ _{45}^{101-109}Rh$ and
positive parity states of the odd-neutron nuclei, $
_{56}^{127-137}Ba$. In order to obtain energy spectrum and
realistic calculation for these nuclei, we need to specify
Hamiltonian parameters (11), (12). Eigenvalues of these systems
are obtained by solving Bethe-Ansatz equations with least square
fitting processes to experimental data to obtain constants of
Hamiltonian. The best fits for Hamiltonian's parameters, namely $
\alpha$ , $ \beta $ , $\delta $ and  $\gamma  $,  used in the
present work are shown in table 1, 2.Tables 3(a,b...e) and 4(a,b...f) are shown calculated energy spectra along with the experimental values .  Figure.8 and Figure.9 are also shown
a comparison between the available experimental levels and the
predictions of our results for the $ _{45}^{101-109}Rh$   and $
_{56}^{127-137}Ba$ isotopes in the low-lying region of spectra.
An acceptable degree of agreement is obvious between them. We
have tried to extract the best set of parameters which reproduce
these complete spectra with minimum variations. It means that our
suggestion to use this transitional Hamiltonian for the
description of the Rh and Ba isotopic chain would not have any
contradiction with other theoretical studies done with special
hypotheses about mixing of intruder and normal configurations. On
the other hand, predictions of our model for the control
parameter of considered nuclei, C, describe the vibrational,
i.e.$ C= 0$, or rotational, namely $C= 1$, confirm this mixing of
both vibrating and rotating structures in these nuclei when
$C\sim 0.5\rightarrow 0.65$. Fig.10 and Fig.11 display a comparison
between the calculated continues energy differences and
experimental data for Ba and Rh Isotopes, respectively. It can be
seen from figs that our results for Ba Isotopes are better than
for Rh Isotopes.

One of the most basic structural predictions of $ U^{BF}
(5)-O^{BF} (6) $ transition is a $\frac
{E(\nu_{d}=2)}{E(\nu_{d}=1)})$ value. The ratio equal to $ 2.2-2.3 $
indicates the spectrum of transitional nuclei \cite{3,4,15,30}. Thus we
calculated this quantity for Rh and Ba Isotopes. Fig.12 shows
$\frac {E(\nu_{d}=2)}{E(\nu_{d}=1)})$ prediction values for Rh and Ba
Isotopes. For Rh isotopes this value evolves from $ 2.1 $ to $
2.8 $ while Ba isotopes vary of $ 2.85 $ to $1.9 $. Fig.12
displays that $\frac {E(\nu_{d}=2)}{E(\nu_{d}=1)})$ values for $
_{45}^{105}Rh $ and $ _{56}^{133-135}Ba $ isotopes are
approximately $2.2-2.4$.
 \begin{table}
\begin{tabular}{p{9.3cm}}

\footnotesize Table 1. \footnotesize Parameters of Hamiltonian
(11) used in the calculation of the Rh Isotopes.
\footnotesize All parameters  are given in keV.\\
\end{tabular}

\begin{tabular}{cccccccc}

\hline
Nucleus      &$ N $    & $C$ & $\alpha$ & $\beta$ & $\delta$ & $\gamma$ & $\sigma$    \\
\hline
$ _{45}^{101}Rh$   &5           &           0.06     &          105.99    &       4.2572  &         1.6782    &        1.2336    &     194.78 \\
$ _{45}^{103}Rh$   &6           &            0.46    &            52       &          -1.053  &           1.5518 &          23.217 &       136.5\\
$ _{45}^{105}Rh$   &7            &           0.54    &           70.61      &       1.6273     &        8.7337   &      -0.0331    &     97.38\\
$ _{45}^{107}Rh$   &8             &          0.65    &           196.30    &        3.5077   &        -11.734    &      7.7931   &      120.48\\
$ _{45}^{109}Rh$   &9              &         0.7      &           245.269   &       2.438     &      -24.072   &     -181.02      &   184.79\\

\hline
\end{tabular}
\end{table}

\begin{table}
\begin{center}
\begin{tabular}{p{9.3cm}}

\footnotesize Table 2. \footnotesize Parameters of Hamiltonian
(11) and (12) used in the calculation of the Ba Isotopes.
\footnotesize All parameters  are given in keV.\\
\end{tabular}

\begin{tabular}{ccccccccc}

\hline
Nucleus      &$ N $    & $C$ & $\alpha$ & $\beta$ &  $\delta$ & $\gamma$ & $\sigma$    \\
\hline
$ _{56}^{127}Ba$   &8             &0.78             &3.46                        &-0.0303        &-1.1481         &19.098              &  62.48 \\
$ _{56}^{129}Ba$   &7              & 0.8           &13.77                     &-2.685         & -8.5776         & 15.69               &     94.25\\
$ _{56}^{131}Ba$   &6              &0.77          &0.578                     &-3.3839          &-3.3319         & 27.12                &  128\\
$ _{56}^{133}Ba$   &5              &0.68           & 21.37                    &-0.0274         & 16.46            &-1.75                 & 119\\
$ _{56}^{135}Ba$   &4             &0.65          &48.69                     & 3.32                  &   -            &    36.57               &  86.84\\
$ _{56}^{137}Ba$   &3             &0.75           &366                     & 0.194                   & -              &  23.047               &128.6\\
\hline
\end{tabular}
\end{center}
\end{table}

\begin{table}
\begin{center}
\begin{tabular}{p{6.3cm}}

\footnotesize Table 3a. \footnotesize .Energy spectra for $  _{45}^{101}Rh $   isotope\\
\end{tabular}

\begin{tabular}{cccccc}

\hline
$  _{45}^{101}Rh $    &$ J^{\pi}$ & \quad $ K $ & \quad $ \nu_{d}$  & \quad $E_{exp} $ & \quad $ E_{cal} $   \\
\hline
\quad & $(1/2)_{1}^{-} $  &  \quad       2&\quad  0 &\quad  0 &\quad  0 \\
\quad & $(3/2)_{1}^{-}  $&  \quad 2&\quad 1 &\quad 305.5& \quad 267.3\\
\quad &$(5/2)_{1}^{-}  $ & \quad 2&\quad 1 &\quad 305.5 &\quad 273.5\\
\quad &$(3/2)_{2}^{-} $ & \quad 1&\quad 2 &\quad 355.3& \quad 665.8\\
\quad &$(5/2)_{2}^{-} $ & \quad 1&\quad 2& \quad 355.3& \quad 671.9\\
\quad &$(7/2)_{1}^{-} $ & \quad 1&\quad 2& \quad 851.4& \quad 704.1\\
\quad &$(9/2)_{1}^{-} $  &\quad 1&\quad 2& \quad 851.4& \quad 715.2\\
\quad &$(9/2)_{2}^{-} $ & \quad 1&\quad 2& \quad 899.3& \quad 821.2\\
\quad &$(5/2)_{3}^{-} $ & \quad 2&\quad 1& \quad 996.4& \quad 794.6\\
\quad &$(3/2)_{4}^{-} $&  \quad 1&\quad 2& \quad 1058& \quad 997.9\\
\quad &$(5/2)_{4}^{-} $ & \quad 1&\quad 2& \quad 1058 &\quad 971.8\\
\quad &$(1/2)_{2}^{-} $  &\quad 2&\quad 0& \quad 1531& \quad 1472.9\\
\hline
\end{tabular}
\end{center}
\end{table}

\begin{table}
\begin{center}
\begin{tabular}{p{6.3cm}}

\footnotesize Table 3b. \footnotesize .Energy spectra for $  _{45}^{103}Rh $   isotope\\
\end{tabular}

\begin{tabular}{cccccc}

\hline
$  _{45}^{103}Rh $    &$ J^{\pi}$ & \quad $ K $ & \quad $ \nu_{d}$  & \quad $E_{exp} $ & \quad $ E_{cal} $   \\
\hline
\quad &   $(1/2)_{1}^{-} $     &   \quad 3 &\quad 0 & \quad 0 & \quad 0  \\
\quad &$(3/2)_{1}^{-} $       & \quad 2&\quad 1& \quad 294.984& \quad 334\\
\quad  & $(5/2)_{1}^{-} $     &   \quad 2&\quad 2& \quad 357.408 &\quad 450.6 \\
 \quad&   $(1/2)_{2}^{-} $   &     \quad 3&\quad 0 &\quad 803.07& \quad 598.6 \\
 \quad& $(3/2)_{2}^{-} $      &  \quad 2&\quad 2 &\quad 803.07& \quad 692.4 \\
 \quad& $(7/2)_{1}^{-} $     &   \quad 2&\quad 2 &\quad 847.58& \quad 705.4 \\
\quad& $(5/2)_{2}^{-} $      &  \quad 2&\quad 2& \quad 880.47& \quad 809 \\
\quad & $(9/2)_{1}^{-} $      &  \quad 2&\quad 2 &\quad 920.1 &\quad 915.3 \\
\quad & $(3/2)_{3}^{-} $      &  \quad 2&\quad 1& \quad 1277.04 &\quad 1226.7 \\
\quad & $(13/2)_{1}^{-} $     &   \quad 1&\quad 3 &\quad 1637.64 &\quad 1576.5 \\
 \quad& $(15/2)_{1}^{-} $      &  \quad 1&\quad 4 &\quad 2221.2 &\quad 2036.3 \\
\quad & $(17/2)_{1}^{-} $      &  \quad 1&\quad 4 &\quad 2345.35 &\quad 2432.6 \\
\quad & $(17/2)_{2}^{-} $      &  \quad 0&\quad 5& \quad 2418.6 &\quad 2065.5 \\

\hline
\end{tabular}
\end{center}
\end{table}

\begin{table}
\begin{center}
\begin{tabular}{p{6.3cm}}

\footnotesize Table 3c. \footnotesize .Energy spectra for $  _{45}^{105}Rh $   isotope\\
\end{tabular}

\begin{tabular}{cccccc}

\hline
$  _{45}^{105}Rh $    &$ J^{\pi}$ & \quad $ K $ & \quad $ \nu_{d}$  & \quad $E_{exp} $ & \quad $ E_{cal} $   \\
\hline
\quad &$(1/2)_{1}^{-} $   &     \quad 3&\quad 0 &\quad 129.781& \quad 129.8\\
\quad &$(3/2)_{1}^{-}  $ & \quad 3&\quad 1& \quad 392.65& \quad 229.2 \\
\quad &$(5/2)_{1}^{-}  $ & \quad 3&\quad 1 &\quad 455.61& \quad 511.7\\
\quad &$(3/2)_{2}^{-} $  &\quad 2&\quad 2 &\quad 762.11& \quad 811.2\\
\quad &$(3/2)_{3}^{-} $  &\quad 3&\quad 1& \quad 783 &\quad 703.6\\
\quad &$(5/2)_{2}^{-} $  &\quad 2&\quad 2 &\quad 817 &\quad 821.7\\
\quad &$(7/2)_{1}^{-} $  &\quad 2&\quad 2& \quad 817 &\quad 833.1\\
\quad &$(5/2)_{3}^{-} $  &\quad 2&\quad 3& \quad 866& \quad 868.2\\
\quad &$(7/2)_{2}^{-} $  &\quad 2&\quad 3& \quad 898& \quad 861.5\\
\quad &$(7/2)_{3}^{-} $  &\quad 2&\quad 3 &\quad 976& \quad 937.9\\
\quad &$(9/2)_{1}^{-} $  &\quad 2&\quad 2 &\quad 976& \quad 1043.5 \\
\quad &$(3/2)_{4}^{-} $  &\quad 2&\quad 2 &\quad 1147& \quad 1312.7\\
\quad &$(5/2)_{4}^{-} $  &\quad 2&\quad 2 &\quad 1147& \quad 1311.9 \\
\hline
\end{tabular}
\end{center}
\end{table}

\begin{table}
\begin{center}
\begin{tabular}{p{6.3cm}}

\footnotesize Table 3d. \footnotesize .Energy spectra for $  _{45}^{107}Rh $   isotope\\
\end{tabular}

\begin{tabular}{cccccc}

\hline
$  _{45}^{107}Rh $    &$ J^{\pi}$ & \quad $ K $ & \quad $ \nu_{d}$  & \quad $E_{exp} $ & \quad $ E_{cal} $   \\
\hline
\quad &$(1/2)_{1}^{-} $&        \quad 4&\quad 0 &\quad 268.36 &\quad 268.4 \\
\quad &$(3/2)_{1}^{-} $  &      \quad 3&\quad 1 &\quad 485.66 &\quad 345.4  \\
\quad &$(5/2)_{1}^{-} $  &      \quad 3&\quad 1 &\quad 543.84 &\quad 410.9\\
\quad & $(9/2)_{1}^{-} $ &       \quad 3&\quad 2 &\quad 559.97 &\quad 423.4 \\
\quad & $(3/2)_{2}^{-} $  &      \quad 3&\quad 2& \quad 752.55 &\quad 818.8\\
\quad &$(5/2)_{2}^{-} $   &     \quad 3&\quad 2 &\quad 877.75& \quad 870.4 \\
\quad &$(3/2)_{3}^{-} $   &     \quad3&\quad 1& \quad 974.44& \quad 921.9\\
\quad &$(5/2)_{3}^{-} $    &    \quad 2&\quad 3 &\quad 974.44& \quad 958.6\\
\quad &$(7/2)_{1}^{-} $ &       \quad 3&\quad 2 &\quad 974.44& \quad 780.9 \\
\quad  &$(3/2)_{4}^{-} $    &    \quad 3&\quad 2 &\quad 1009.76& \quad 1016.3\\
\quad & $(5/2)_{4}^{-} $   &     \quad 3&\quad 2 &\quad 1009.76 &\quad 1055.3 \\
\quad & $(7/2)_{2}^{-} $  &      \quad 2&\quad 3 &\quad 1251 &\quad 1099.5\\
\quad &$(1/2)_{2}^{-} $    &    \quad 4&\quad 0 &\quad 1334& \quad 1431.2  \\
\hline
\end{tabular}
\end{center}
\end{table}

\begin{table}
\begin{center}
\begin{tabular}{p{6.3cm}}

\footnotesize Table 3e. \footnotesize .Energy spectra for $  _{45}^{109}Rh $   isotope\\
\end{tabular}

\begin{tabular}{cccccc}

\hline
$  _{45}^{109}Rh $    &$ J^{\pi}$ & \quad $ K $ & \quad $ \nu_{d}$  & \quad $E_{exp} $ & \quad $ E_{cal} $   \\
\hline
 \quad&    $(1/2)_{1}^{-} $ &       \quad 4&\quad 0 &\quad 374.1& \quad 374.1  \\
\quad&  $(3/2)_{1}^{-} $    &    \quad 4&\quad 1 &\quad 568.2& \quad 436.6\\
 \quad  & $(5/2)_{1}^{-} $  &      \quad 4&\quad 1 &\quad 623.2& \quad 401.5 \\
 \quad &  $(3/2)_{2}^{-} $  &      \quad 3&\quad 2& \quad 704.9& \quad 758.1 \\
\quad & $(5/2)_{2}^{-} $    &    \quad 3&\quad 2 &\quad 856.1 &\quad 819.6 \\
\quad& $(5/2)_{3}^{-} $     &   \quad 3&\quad 3& \quad 926.9& \quad 827.6 \\
\quad & $(3/2)_{3}^{-} $    &    \quad 4&\quad 1 &\quad 1162.3 &\quad 1378.8 \\
\quad & $(3/2)_{4}^{-} $    &    \quad 3&\quad 2 &\quad 1214.3& \quad 1003.3 \\
\quad& $(5/2)_{4}^{-} $     &   \quad 3&\quad 2 &\quad 1283.9 &\quad 985.6 \\
\quad & $(1/2)_{2}^{-} $    &    \quad 4&\quad 0& \quad 1631& \quad 1876.6 \\
 \quad&$(3/2)_{5}^{-} $     &   \quad 4&\quad 1 &\quad 1631 &\quad 1627.4 \\
 \quad &$(1/2)_{3}^{-} $    &    \quad 4&\quad 0& \quad 1753& \quad 1703 \\
\quad& $(3/2)_{6}^{-} $     &   \quad 4&\quad 1& \quad 1753& \quad 1750 \\
\hline
\end{tabular}
\end{center}
\end{table}

\begin{table}
\begin{center}
\begin{tabular}{p{6.3cm}}

\footnotesize Table 4a. \footnotesize .Energy spectra for $ _{56}^{127}Ba $   isotope\\
\end{tabular}

\begin{tabular}{cccccc}

\hline
$ _{56}^{127}Ba $    &$ J^{\pi}$ & \quad $ K $ & \quad $ \nu_{d}$  & \quad $E_{exp} $ & \quad $ E_{cal} $   \\
\hline
 \quad & $(1/2)_{1}^{+} $  &      \quad 4&\quad 0& \quad 0&\quad 0\\
 \quad& $(3/2)_{1}^{+}  $ & \quad 3&\quad 1 &\quad 56.1 &\quad 51.4\\
  \quad&$(5/2)_{1}^{+}  $ & \quad 3&\quad 1 &\quad 81 &\quad 146.9\\
  \quad&$(7/2)_{1}^{+} $  &\quad 3&\quad 2 &\quad 195.1 &\quad 265.7\\
  \quad&$(3/2)_{2}^{+} $  &\quad 3&\quad 2 &\quad 269.5 &\quad 256.3\\
  \quad&$(5/2)_{2}^{+} $  &\quad 3&\quad 2 &\quad 269.5 &\quad 237.2\\
 \quad& $(7/2)_{1}^{+} $  &\quad 2&\quad 3 &\quad 324.1 &\quad 376.3\\
 \quad& $(7/2)_{2}^{+} $ & \quad 2&\quad 3 &\quad 374.9 &\quad 367.1\\
 \quad& $(9/2)_{1}^{+} $  &\quad 3&\quad 2 &\quad 415.6&\quad 526.1\\
 \quad& $(11/2)_{1}^{+} $ & \quad 2&\quad 3& \quad 668.9 &\quad 723.8\\
 \quad& $(11/2)_{2}^{+} $ & \quad 2&\quad 4& \quad 867.9 &\quad 759.4\\
 \quad& $(13/2)_{1}^{+} $ & \quad 2&\quad 3& \quad 963.6 &\quad 972.1\\
 \quad&$(15/2)_{1}^{+} $  &\quad 2&\quad 4& \quad 1291.2 &\quad 1259.8\\
 \quad&$(15/2)_{2}^{+} $  &\quad 1&\quad 5& \quad 1519.6 &\quad 1425.5\\
 \quad&$(17/2)_{1}^{+} $  &\quad 2&\quad 4& \quad 1654.4 &\quad 1584.4\\
\hline
\end{tabular}
\end{center}
\end{table}

\begin{table}
\begin{center}
\begin{tabular}{p{6.3cm}}

\footnotesize Table 4b. \footnotesize .Energy spectra for$ _{56}^{129}Ba $  isotope\\
\end{tabular}

\begin{tabular}{cccccc}

\hline
$ _{56}^{129}Ba $    &$ J^{\pi}$ & \quad $ K $ & \quad $ \nu_{d}$  & \quad $E_{exp} $ & \quad $ E_{cal} $   \\
\hline
\quad&    $(1/2)_{1}^{+} $   &     \quad 3&\quad 0& \quad 0& \quad 0\\
 \quad&  $(7/2)_{1}^{+} $   &     \quad 2&\quad 2& \quad 8.42& \quad 48.71\\
 \quad & $(3/2)_{1}^{+} $   &     \quad 3&\quad 1& \quad 110.56 &\quad 185.67\\
\quad&   $(3/2)_{2}^{+} $   &     \quad 2&\quad 2& \quad 253.77 &\quad 210.76\\
\quad & $(9/2)_{1}^{+} $    &    \quad 2&\quad 2& \quad 263.1& \quad 249.9\\
 \quad& $(1/2)_{2}^{+} $      &  \quad 3&\quad 0 &\quad 278.58 &\quad 331.88 \\
\quad& $(3/2)_{3}^{+} $       & \quad 3&\quad 1& \quad 278.58& \quad 228.15\\
\quad& $(5/2)_{1}^{+} $      &  \quad 3&\quad 1 &\quad 278.58 &\quad 292.83\\
\quad& $(5/2)_{2}^{+} $      &  \quad 2&\quad 2 &\quad 318.4 &\quad 290.66\\
\quad & $(3/2)_{4}^{+} $      &  \quad 2&\quad 2 &\quad 457.03& \quad 483.53\\
\quad& $(3/2)_{5}^{+} $       & \quad 3&\quad 1 &\quad 459.29& \quad 492.44\\
\quad& $(5/2)_{3}^{+} $      &  \quad 2&\quad 3 &\quad 459.29 &\quad 548.4\\
\quad& $(7/2)_{2}^{+} $      &  \quad 2&\quad 3 &\quad 467.3& \quad 589.6\\
\quad& $(11/2)_{1}^{+} $      &  \quad 2&\quad 3 &\quad 544.7& \quad 441.257\\
\quad& $(13/2)_{1}^{+} $       & \quad 2&\quad 3 &\quad 864.1 &\quad 918.66\\
\hline
\end{tabular}
\end{center}
\end{table}
\begin{table}
\begin{center}
\begin{tabular}{p{6.3cm}}

\footnotesize Table 4c. \footnotesize .Energy spectra for$ _{56}^{131}Ba $  isotope\\
\end{tabular}

\begin{tabular}{cccccc}

\hline
$ _{56}^{131}Ba $    &$ J^{\pi}$ & \quad $ K $ & \quad $ \nu_{d}$  & \quad $E_{exp} $ & \quad $ E_{cal} $   \\
\hline
 \quad  &    $(1/2)_{1}^{+} $  &      \quad 3&\quad 0& \quad 0& \quad 0 \\
 \quad &    $(3/2)_{1}^{+} $      &  \quad 2&\quad 1 &\quad 108.08 &\quad 70.5 \\
 \quad&    $(3/2)_{2}^{+} $   &     \quad 2&\quad 2 &\quad 285.25 &\quad 359.2  \\
 \quad &    $(5/2)_{1}^{+} $     &   \quad 2&\quad 1& \quad 316.587 &\quad 424.7  \\
\quad  &    $(1/2)_{2}^{+} $  &      \quad 3&\quad 0& \quad 365.16& \quad 487.8 \\
\quad  &    $(3/2)_{3}^{+} $      &  \quad 2&\quad 1 &\quad 525.85 &\quad 577.3 \\
 \quad  &    $(5/2)_{2}^{+} $    &    \quad 2&\quad 2& \quad 525.85& \quad 494.8 \\
 \quad &    $(7/2)_{1}^{+} $     &   \quad 2&\quad 2 &\quad 543.11 &\quad 638.1  \\
 \quad &    $(3/2)_{4}^{+} $     &   \quad 2&\quad 1& \quad 561.752 &\quad 643.5  \\
\quad &    $(5/2)_{3}^{+} $      &  \quad 1&\quad 3& \quad 561.75 &\quad 473.3  \\
\quad   &    $(3/2)_{5}^{+} $    &    \quad 0&\quad 5& \quad 718.78 &\quad 573.7 \\
\quad  &    $(5/2)_{4}^{+} $     &   \quad 2&\quad 1& \quad 718.78& \quad 778.6  \\
\quad  &    $(1/2)_{2}^{+} $     &   \quad 1&\quad 3 &\quad 719.49& \quad 698.1  \\
 \quad&    $(15/2)_{1}^{+} $     &   \quad 1&\quad 4& \quad 1796.4& \quad 1821.6  \\
 \quad&    $(17/2)_{1}^{+} $     &   \quad 1&\quad 4& \quad 2121.7 &\quad 2282.6  \\
\hline
\end{tabular}
\end{center}
\end{table}

\begin{table}
\begin{center}
\begin{tabular}{p{6.3cm}}

\footnotesize Table 4d. \footnotesize .Energy spectra for$ _{56}^{133}Ba $  isotope\\
\end{tabular}

\begin{tabular}{cccccc}

\hline
$ _{56}^{133}Ba $    &$ J^{\pi}$ & \quad $ K $ & \quad $ \nu_{d}$  & \quad $E_{exp} $ & \quad $ E_{cal} $   \\
\hline

  \quad  &    $(1/2)_{1}^{+} $  &      \quad 2&\quad 0& \quad 0& \quad 0 \\
 \quad &    $(3/2)_{1}^{+} $      &  \quad 2&\quad 1 &\quad 12.3 &\quad 64.86 \\
 \quad&    $(5/2)_{1}^{+} $   &     \quad 2&\quad 1 &\quad 291.2 &\quad 175.93  \\
\quad  &    $(3/2)_{2}^{+} $  &      \quad 1&\quad 2& \quad 302.4& \quad 419.34 \\
\quad  &    $(1/2)_{2}^{+} $      &  \quad 1&\quad 2 &\quad 539.8 &\quad 443.23 \\
 \quad  &    $(7/2)_{1}^{+} $    &    \quad 1&\quad 2& \quad 577.5& \quad 701.22 \\
 \quad &    $(3/2)_{3}^{+} $     &   \quad 1&\quad 3 &\quad630.6
 &\quad 705.24  \\
  \quad &    $(5/2)_{2}^{+} $     &   \quad 1&\quad 2& \quad 630.6 &\quad 720.84  \\
  \quad &    $(5/2)_{3}^{+} $     &   \quad 1&\quad 3& \quad 676.5 &\quad 675.45  \\
\quad &    $(3/2)_{4}^{+} $      &  \quad 0&\quad 5& \quad 676.5 &\quad 726.015  \\
\quad   &    $(5/2)_{4}^{+} $    &    \quad 1&\quad 3& \quad 858.5 &\quad 696.45 \\
\quad  &    $(9/2)_{1}^{+} $     &   \quad 1&\quad 2& \quad 883.3& \quad 933.86  \\
\quad  &    $(7/2)_{2}^{+} $     &   \quad 1&\quad 3 &\quad 1112.3& \quad938.98  \\

\hline
\end{tabular}
\end{center}
\end{table}

\begin{table}
\begin{center}
\begin{tabular}{p{6.3cm}}

\footnotesize Table 4e. \footnotesize .Energy spectra for$ _{56}^{135}Ba $  isotope\\
\end{tabular}

\begin{tabular}{cccccc}

\hline
$ _{56}^{135}Ba $    &$ J^{\pi}$ & \quad $ K $ & \quad $ \nu_{d}$  & \quad $E_{exp} $ & \quad $ E_{cal} $   \\
\hline

  \quad  &    $(3/2)_{1}^{+} $  &      \quad 2&\quad 0& \quad 0& \quad 0 \\
 \quad &    $(1/2)_{1}^{+} $      &  \quad 1&\quad 1 &\quad 220.954 &\quad 200.66 \\
 \quad&    $(5/2)_{1}^{+} $   &     \quad 1&\quad 1 &\quad 480.52 &\quad 502.31  \\
 \quad &    $(3/2)_{2}^{+} $     &   \quad 1&\quad 1& \quad 587.82 &\quad 696.93  \\
\quad  &    $(3/2)_{3}^{+} $  &      \quad 2&\quad 0& \quad 855& \quad 866.9 \\
\quad  &    $(1/2)_{2}^{+} $      &  \quad 1&\quad 2 &\quad 910.35 &\quad 985.569 \\
 \quad  &    $(5/2)_{2}^{+} $    &    \quad 1&\quad 2& \quad 979.966& \quad 904.815 \\
 \quad &    $(3/2)_{4}^{+} $     &   \quad 1&\quad 2 &\quad 979.969 &\quad 1145.37\\
\hline
\end{tabular}
\end{center}
\end{table}

\begin{figure}[htb]
\begin{center}
\includegraphics[height=6cm]{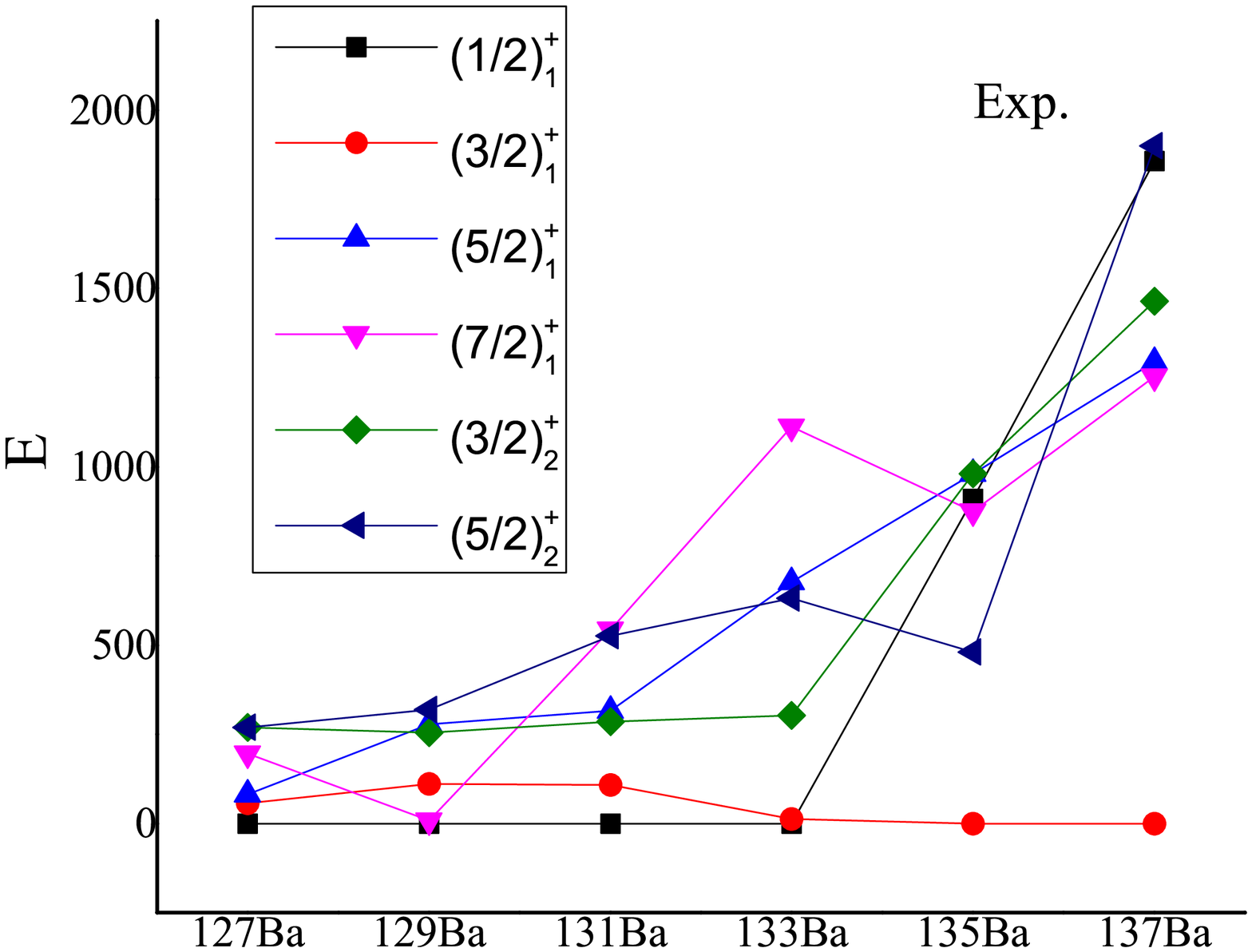}
\includegraphics[height=6cm]{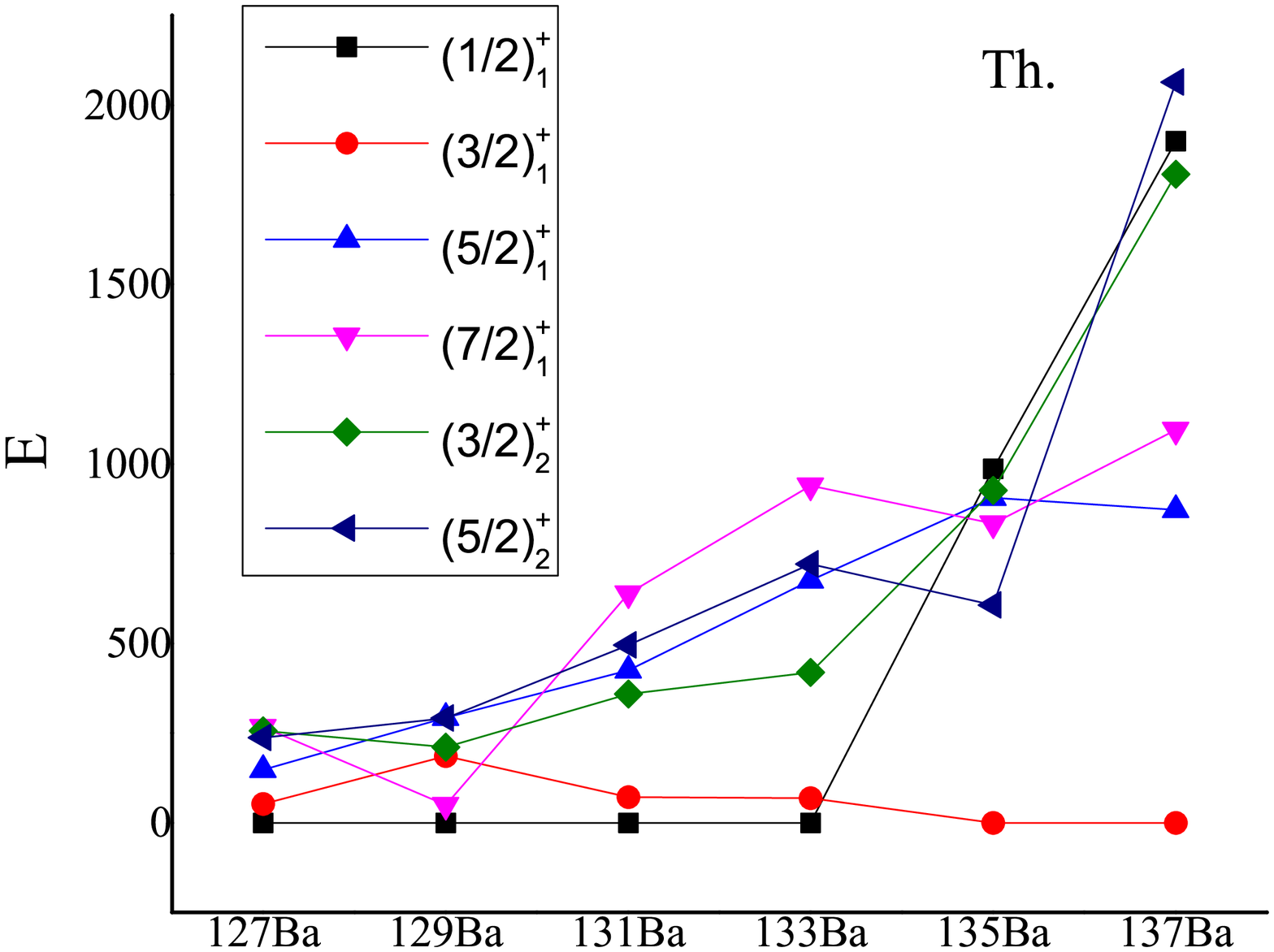}
\caption{Comparison between calculated and experimental spectra
of positive parity states in Ba Isotopes. The parameters of the
calculation are given in Tables 2. In the experimental spectra,
taken from \cite{21}.\label{fig:6}}
\end{center}
\end{figure}

\begin{figure}[htb]
\begin{center}
\includegraphics[height=6cm]{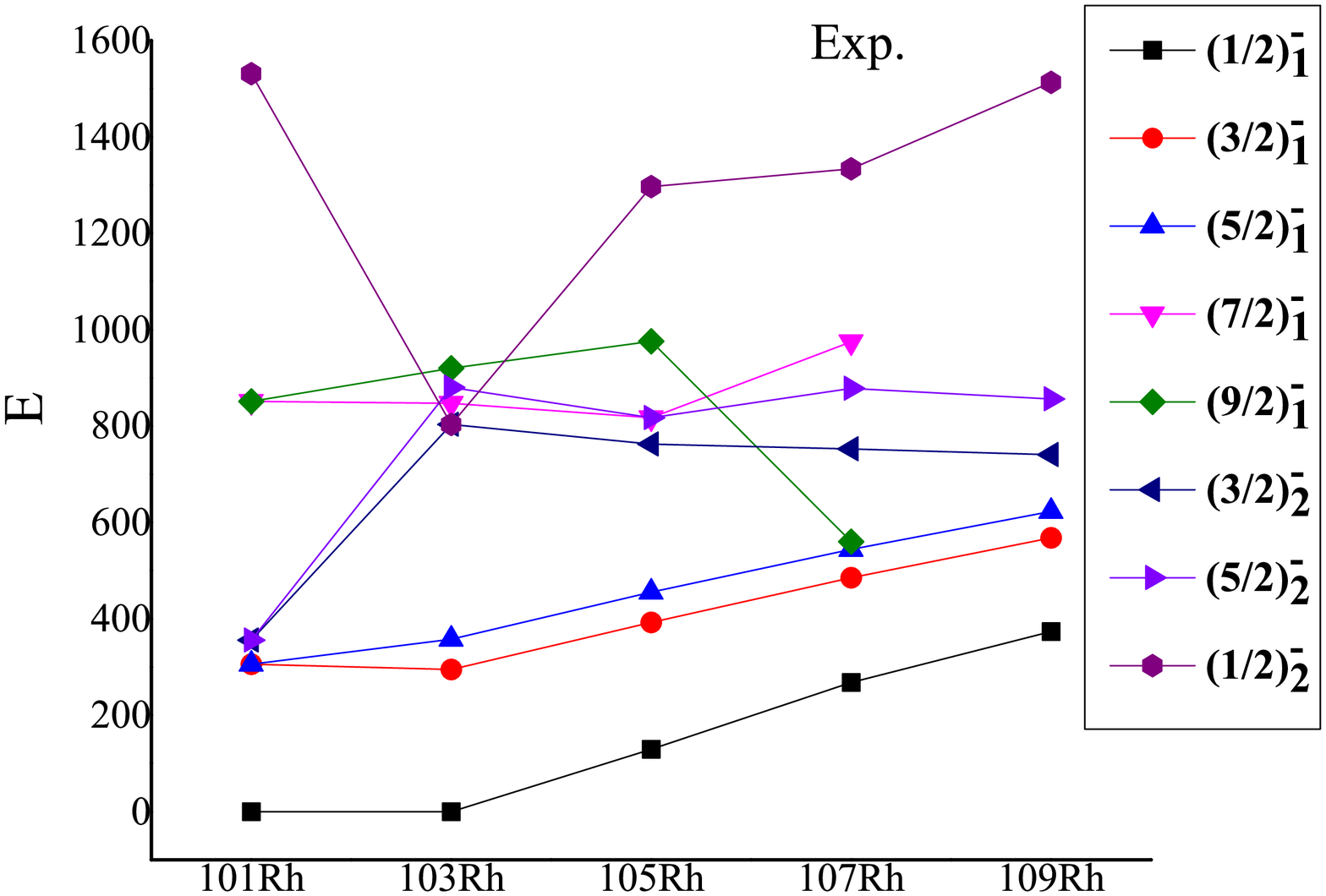}
\includegraphics[height=6cm]{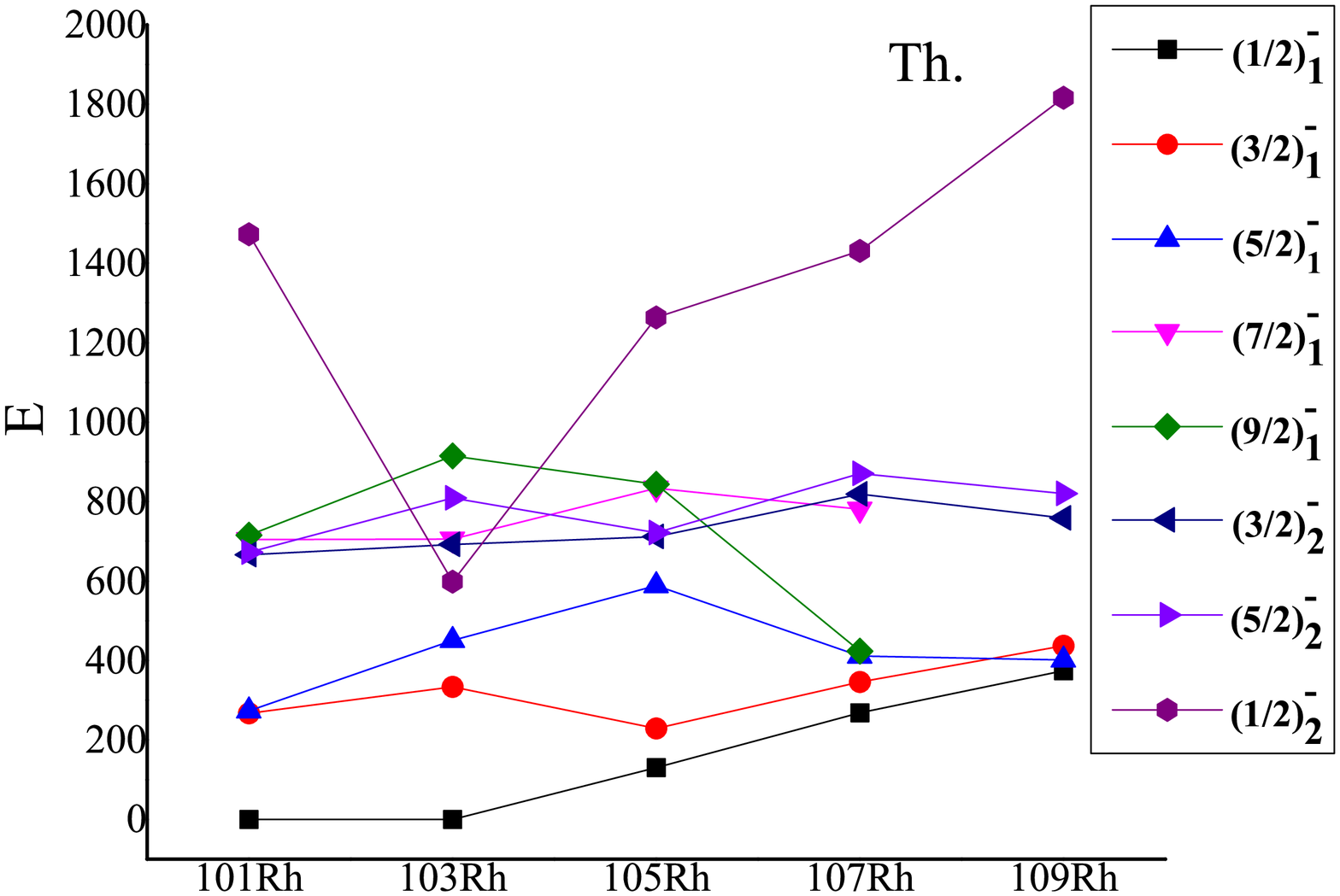}
\caption{Comparison between calculated and experimental spectra
of negative parity states in Rh Isotopes. The parameters of the
calculation are given in Tables 1. In the experimental spectra,
taken from \cite{21}.\label{fig:7}}
\end{center}
\end{figure}

\begin{figure}[htb]
\begin{center}
\includegraphics[height=6cm]{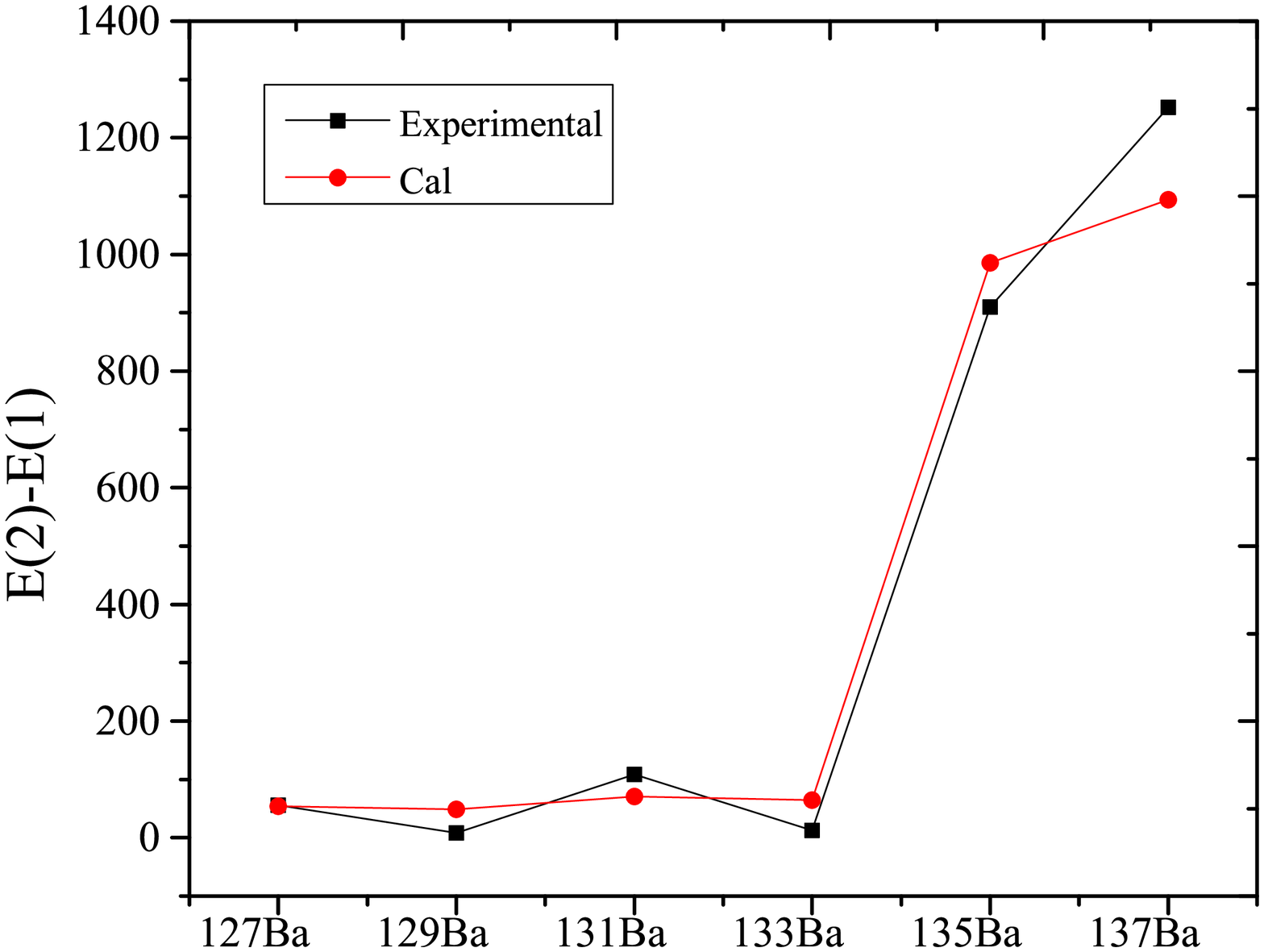}
\includegraphics[height=6cm]{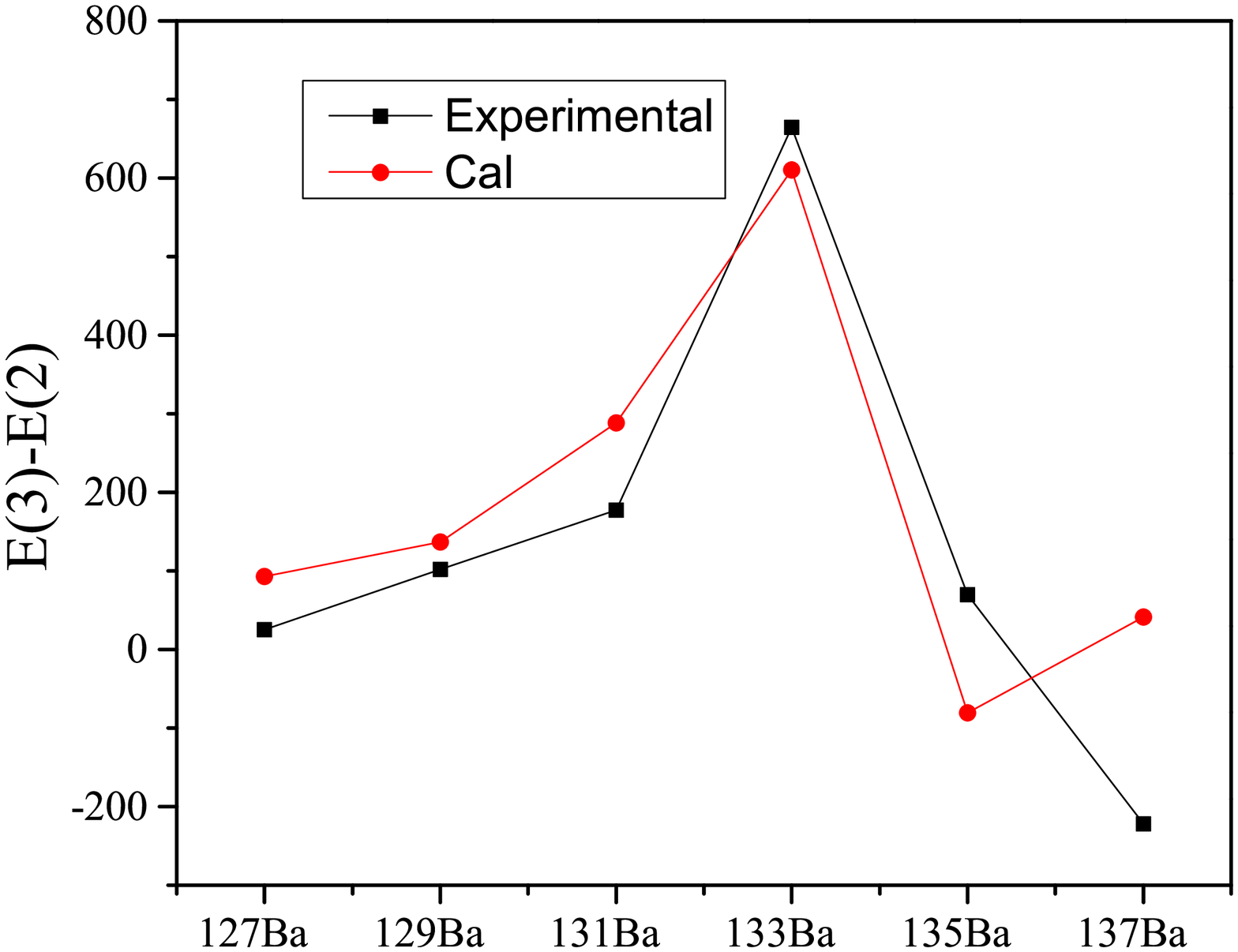}
\caption{A comparison between the calculated continues energy
differences and experimental data for Ba. In the experimental
spectra, taken from \cite{21}.\label{fig:8}}
\end{center}
\end{figure}

\begin{figure}[htb]
\begin{center}
\includegraphics[height=6cm]{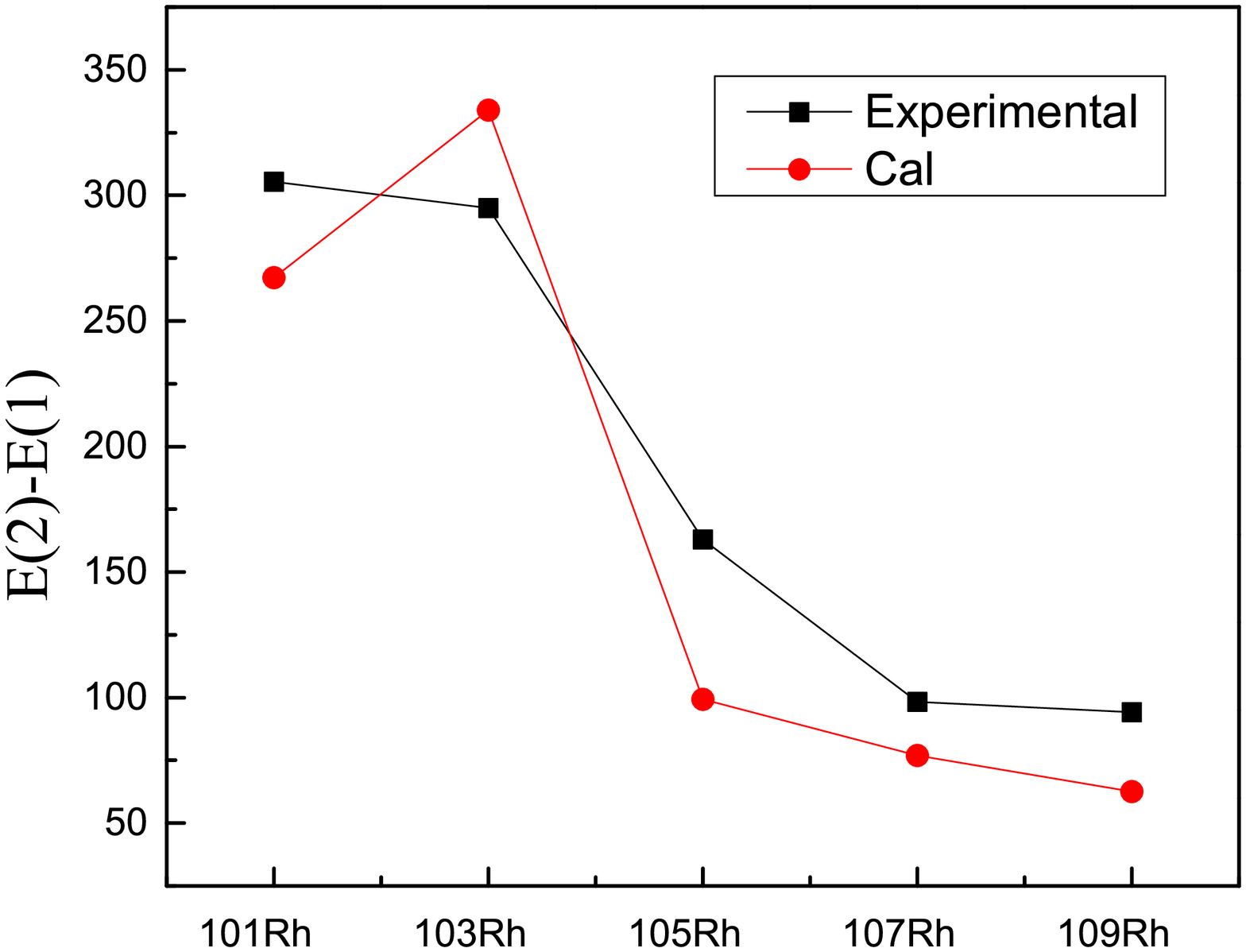}
\includegraphics[height=6cm]{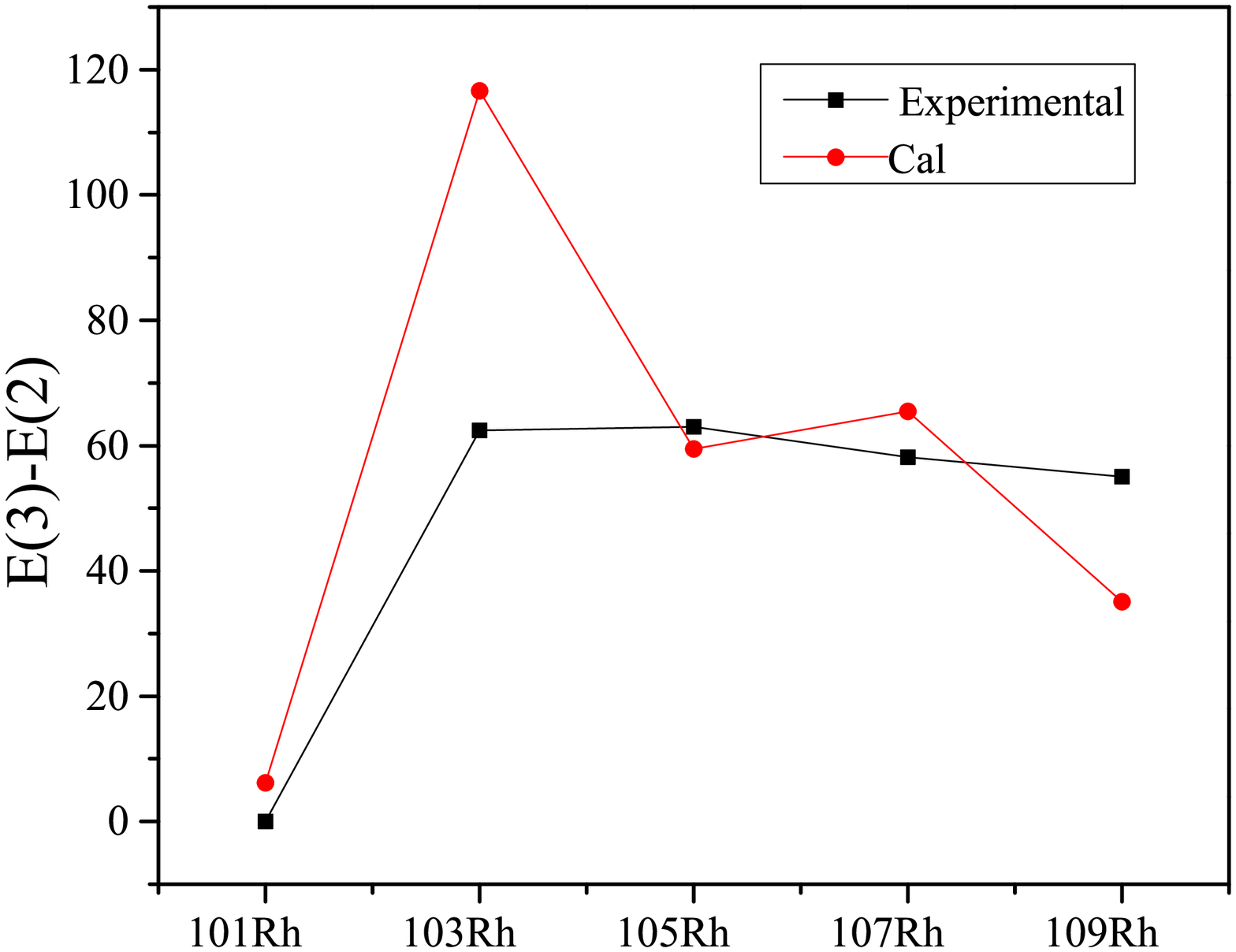}
\caption{A comparison between the calculated continues energy
differences and experimental data for Rh. In the experimental
spectra, taken from \cite{21}.\label{fig:9}}
\end{center}
\end{figure}

\begin{figure}[htb]
\begin{center}
\includegraphics[height=6cm]{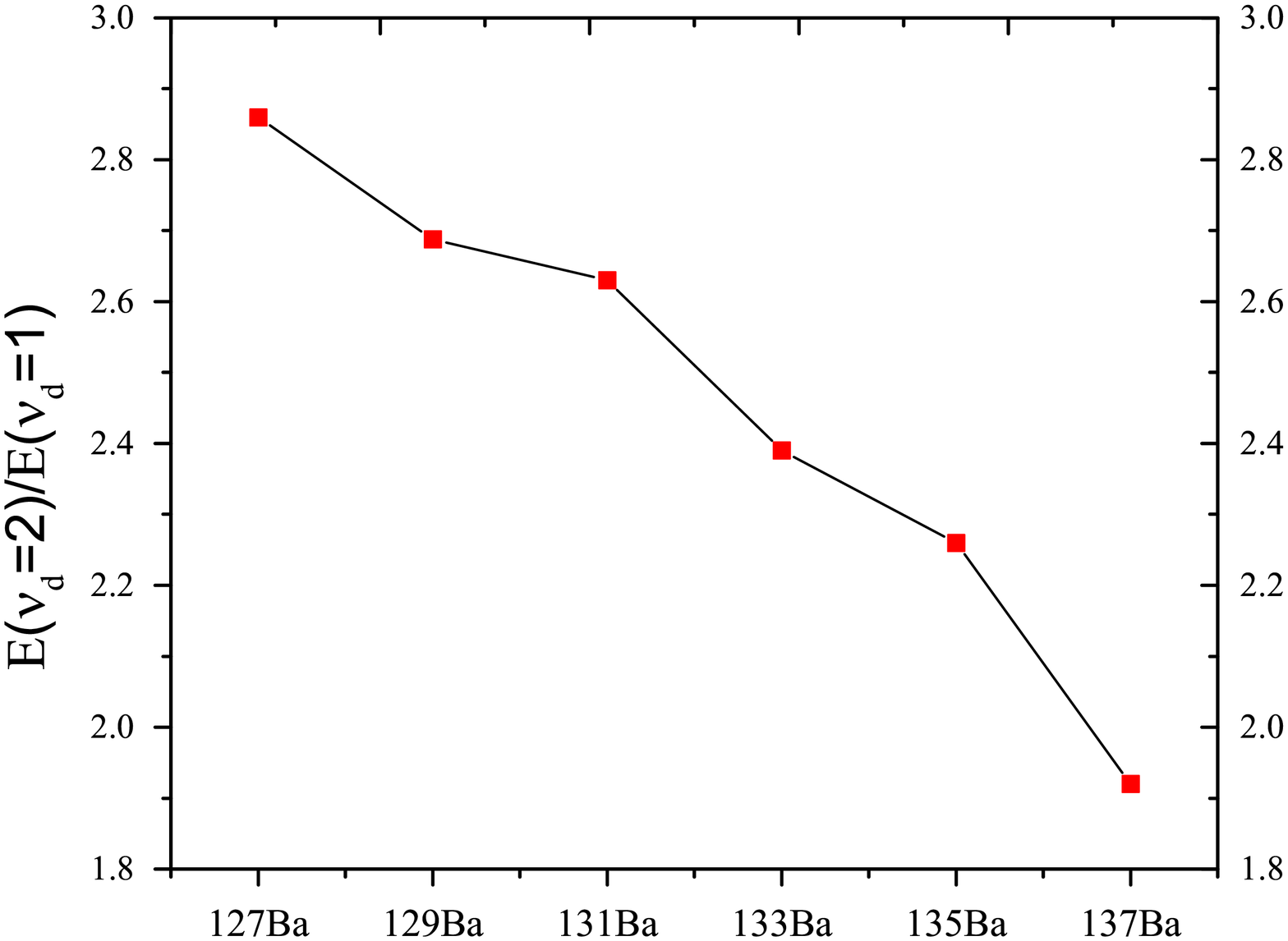}
\includegraphics[height=6cm]{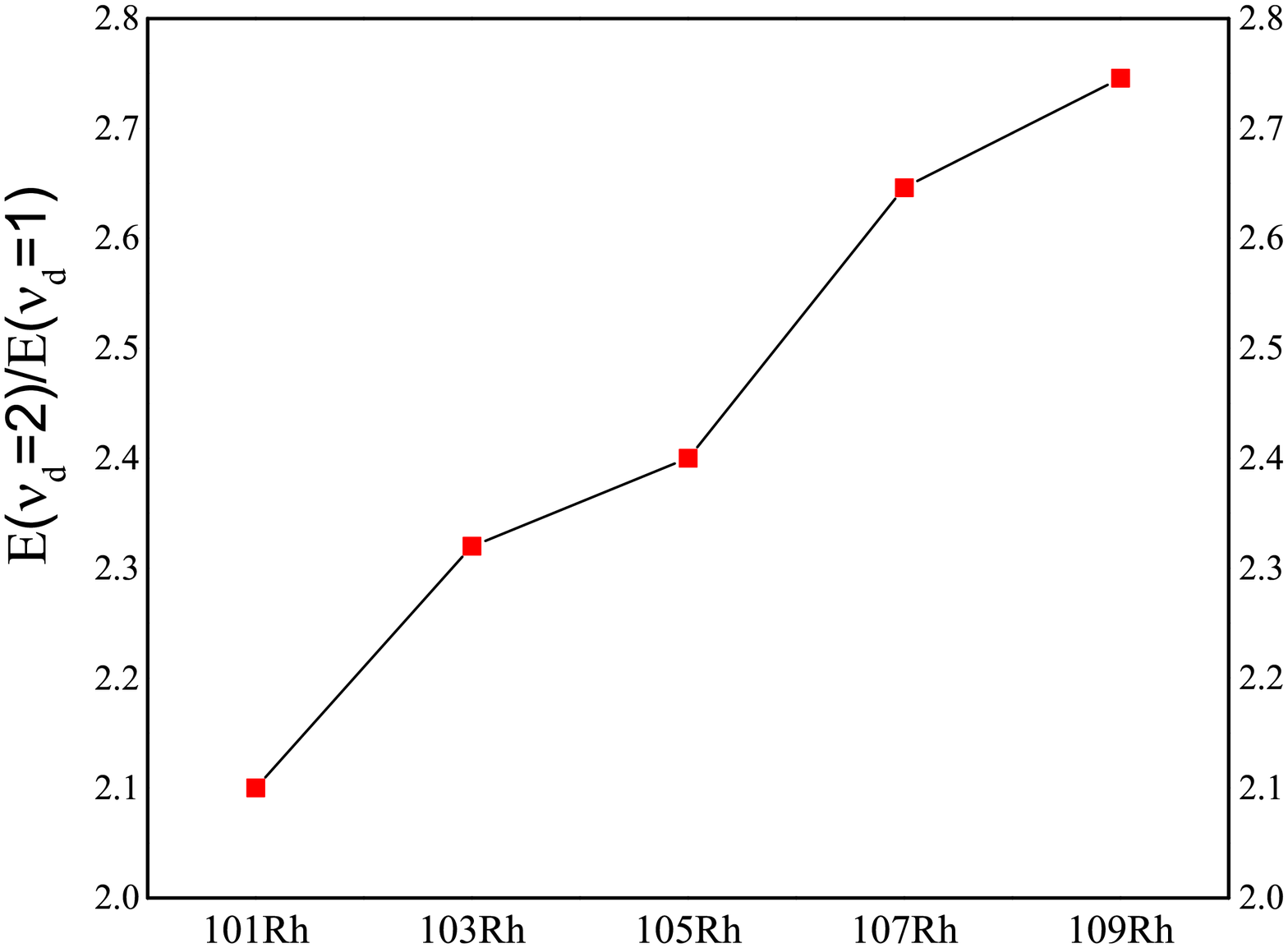}
\caption{$(E( \nu_{d}=2))/(E(\nu_{d}=1))$ prediction values for Rh and Ba
Isotopes. In the experimental spectra, taken from \cite{21}
\label{fig:10}}
\end{center}
\end{figure}

\subsection{$ B(E2) $ transition }
The observables such as electric quadrupole transition
probabilities, $B(E2)$, as well as quadrupole moment ratios
within the low-lying state provide important information about
$QPTs$. In this section we discuss the calculation of E2
transition strengths and compare the results with the available
experimental. The electric quadrupole transition operator $\hat
{T}^{(E2)}$ in odd-A nuclei consists of a bosonic and a fermionic
part\cite{8,25}:
\begin{equation}
\hat {T}^{(E2)} =\hat {T} _{B}^{(E2)} +\hat {T} _{F}^{(E2)}
\end{equation}
With
\begin{equation}
{T} _{B, \mu}^{(E2)} =q_{2} [s^{+}\times \tilde{d} + d^{+} \times
\tilde{s}]_{\mu}^{(2) }+q'_{2} [d^{+} \times
\tilde{d}]_{\mu}^{(2) }=q_{B} Q_{B,\mu}
\end{equation}
\begin{equation}
Q_{B,\mu}=[s^{+}\times \tilde{d} + d^{+} \times
\tilde{s}]_{\mu}^{(2) }+\chi [d^{+} \times \tilde{d}]_{\mu}^{(2) }
\end{equation}
\begin{equation}
 {T} _{F}^{(E2)} =q_{f} \sum_{jj'}Q_{jj'}[a_{j}^{+}\times \tilde{a}_{j'} ]^{(2) }
\end{equation}
Where $ Q_{B} $   and $ Q_{jj'} $   are the boson and fermion
quadrupole operator and $q_{B}$ and $q_{f}$are the effective
boson and fermion charges\cite{8,25}.In the $j=1/2 $ case, the
E2 transitions are completely determined by the bosonic part of
the E2 operator . The bosonic part  have the specific selection
rules, where for former term $ \Delta \nu_{d}=\pm 1 $, $ |\Delta
L |\leq 2 $ and for latter $ \Delta \nu_{d}=0,\pm 2 $, $ |\Delta
L |\leq 0,4 $. The reduced electric quadrupole transition rate
between the $ J_{i}\rightarrow J_{f} $ states is given by
\cite{8}
\begin{equation}
B(E2; \alpha _{i} J_{i} \rightarrow \alpha _{f} J_{f})=\frac
{|\langle \alpha _{f} J_{f} || T^{(E2)} || \alpha _{i}
J_{i}\rangle |^{2} }{2J_{i}+1}
\end{equation}
The electric quadrupole moment for a state with spin $J$ is given
by \cite{8}
\begin{equation}
Q_{J}=\sqrt{\frac {16\pi}{5}} (\frac {J(2J-1)}
{(2J+1)(J+1)(2J+3)})^{\frac {1} {2}}{\langle J || T^{(E2)} ||
J\rangle  }
\end{equation}
For evaluating B(E2), We consider eigenstates (14) that the
normalization factor obtain as:
\begin{equation}
N=\sqrt{{\frac {1}{ \prod_{p=1}^{k} \sum_{i=p}^{k}(\frac
{2C^{2}(k-p+\frac {1}{2}(\nu_s+\frac
{1}{2}))}{(1-C^{2}y_{k+1-p})(1-C^{2}y_{i})}+\frac {2(k-p+\frac
{1}{2}(\nu_d+\frac {5}{2}))}{(1-y_{k+1-p})(1-y_{i})})}}}
\end{equation}
Unfortunately there are very few experimental data available on
electromagnetic properties for odd -mass Ba \cite{29} and Rh
\cite{25} isotopes. The values of effective charge
$(q_{B},q_{f})$ are tabulated in Table.5. Table.6 shows
experimental and calculated values for B(E2) for negative parity
states of $ _{45}^{103}Rh $ , and positive parity states of  $
_{56}^{135}Ba $. The quadrupole moments for $ _{45}^{103}Rh $ and
$ _{56}^{135}Ba $ also display in table.7.  Table.6 and Table.7
show that in general there are a good agreement between the
calculated B(E2) values and the quadrupole moments with the
experimentally data.
 The values of the control parameter, C,
suggest structural changes in nuclear deformation and shape-phase
transitions in Odd-proton Rh isotopes and Odd-neutron Ba
isotopes. Because of the effect of single nucleon on the
transition and especially critical point, exact selection of the
critical point is difficult. With considering this problem, we
proposed $C \sim 0.5- 0.65$ as critical point. So, we conclude
from the values of control parameter which has been obtained,
$\frac {E(\nu_{d}=2)}{E(\nu_{d}=1)})$ value and energy differences
that $ _{45}^{105}Rh $ and  $ _{56}^{133-135}Ba $ isotopes are as
the best candidates for $ U^{BF} (5)-O^{BF} (6)$ transition.
\begin{table}
\begin{center}
\begin{tabular}{p{6.3cm}}

\footnotesize Table 4f. \footnotesize .Energy spectra for$ _{56}^{137}Ba $  isotope\\
\end{tabular}

\begin{tabular}{cccccc}

\hline
$ _{56}^{137}Ba $    &$ J^{\pi}$ & \quad $ K $ & \quad $\nu_{d}$  & \quad $E_{exp} $ & \quad $ E_{cal} $   \\
\hline

  \quad  &    $(3/2)_{1}^{+} $  &      \quad 2&\quad 0& \quad 0& \quad 0 \\
 \quad &    $(7/2)_{1}^{+} $      &  \quad 1&\quad 1 &\quad 1252.5 &\quad 1093.9 \\
 \quad&    $(5/2)_{1}^{+} $   &     \quad 1&\quad 1 &\quad 1294 &\quad 1124.25  \\
 \quad &    $(3/2)_{2}^{+} $     &   \quad 1&\quad 1& \quad 1463.8 &\quad 1607  \\
\quad  &    $(1/2)_{1}^{+} $  &      \quad 1&\quad 2& \quad 1857& \quad 1782.9 \\
\quad  &    $(1/2)_{2}^{+} $      &  \quad 1&\quad 1 &\quad 1836.2 &\quad 1899.2 \\
 \quad  &    $(3/2)_{3}^{+} $    &    \quad 2&\quad 0& \quad 2041& \quad 1940.4 \\
 \quad &    $(5/2)_{2}^{+} $     &   \quad 1&\quad 2 &\quad 1899 &\quad 2063.2\\
 \quad  &    $(3/2)_{4}^{+} $  &      \quad 1&\quad 2& \quad 1899 &\quad 1883.5 \\
  \quad &    $(5/2)_{3}^{+} $      &  \quad 1&\quad 2 &\quad 2041 &\quad 2041.8 \\
  \quad&    $(7/2)_{2}^{+} $   &     \quad 1&\quad 2 &\quad 2230 &\quad 2258  \\
  \quad &    $(9/2)_{1}^{+} $     &   \quad 1&\quad 2& \quad 2230 &\quad 2248.6  \\
 \quad  &    $(7/2)_{3}^{+} $  &      \quad 1&\quad 2& \quad 2340& \quad 2263.5 \\
 \quad  &    $(7/2)_{4}^{+} $  &      \quad 1&\quad 1& \quad 2423.8& \quad 2530 \\
\hline
\end{tabular}
\end{center}
\end{table}
\begin{table}
\begin{center}
\begin{tabular}{p{4cm}}

\footnotesize Table 5.\footnotesize The coefficients of $T
(E_{2} ) $used in the present work
\footnotesize for $ _{45}^{103}Rh$ Rh and $ _{56}^{135}Ba$  isotopes.\\
\end{tabular}

\begin{tabular}{ccc}

\hline
Nucleus      &$ q_{B}(eb) $    &$ q_{f}(eb) $    \\
\hline
$ _{45}^{103}Rh$ & 0.461        &                  0  \\
$ _{56}^{135}Ba$  & 4.7329      &            -0.7194\\

\hline
\end{tabular}
\end{center}
\end{table}

\begin{table}
\begin{center}
\begin{tabular}{p{9.2cm}}
\footnotesize Table 6. \footnotesize B(E2)values for $
_{45}^{103}Rh$ and $ _{56}^{135}Ba$ isotopes. The experimental
data for $ _{45}^{103}Rh$
\footnotesize isotope are taken from \cite{21,25}.The experimental data for $ _{56}^{135}Ba$
isotope are taken from \cite{21,31}.\\
\end{tabular}

\begin{tabular}{ccc}

\hline
Nucleus      &$ J_{i}^{\pi}\longrightarrow J_{j}^{\pi}  $    &$ \frac {B(E_{2} (e^{2} b^{2})}{exp. \quad\quad\quad calc.} $    \\
\hline
$ _{45}^{103}Rh$   &   $(3/2)_{1}^{-} \longrightarrow (1/2)_{1}^{-}$        &    0.109      \quad\quad     0.1172  \\
 \quad\quad &  $(5/2)_{1}^{-} \longrightarrow (1/2)_{1}^{-}$  &0.118      \quad\quad     0.1172\\
\quad\quad &  $(5/2)_{2}^{-} \longrightarrow (1/2)_{1}^{-}$  &0.0044              \quad\quad      0.0044 \\
\quad\quad &  $(5/2)_{2}^{-} \longrightarrow (3/2)_{1}^{-}$  &0.0768               \quad\quad     0.0645 \\
\quad\quad &  $(5/2)_{2}^{-} \longrightarrow (5/2)_{1}^{-}$  &0.015                 \quad\quad    0.0097\\
\quad\quad &  $(7/2)_{1}^{-} \longrightarrow (3/2)_{1}^{-}$  &0.13                  \quad\quad      0.1165\\
\quad\quad &  $(9/2)_{1}^{-} \longrightarrow (5/2)_{1}^{-}$  &0.179                \quad\quad      0.1349 \\
$ _{56}^{135}Ba$   &   $(1/2)_{1}^{+} \longrightarrow (3/2)_{1}^{+}$        &   4.6                     \quad\quad    3.696   \\
 \quad\quad &  $(5/2)_{1}^{+} \longrightarrow (1/2)_{1}^{+}$  &2.65                     \quad\quad     2.913\\
\quad\quad &  $(7/2)_{1}^{+} \longrightarrow (3/2)_{1}^{+}$  & 19.9                             \quad\quad      14.784 \\
\quad\quad &  $(1/2)_{2}^{+} \longrightarrow (3/2)_{1}^{+}$  &11.7                            \quad\quad     14.403 \\
\quad\quad &  $(3/2)_{2}^{+} \longrightarrow (3/2)_{1}^{+}$  &18                                 \quad\quad   14.785 \\
\quad\quad &  $(3/2)_{3}^{+} \longrightarrow (3/2)_{1}^{+}$  &7                                    \quad\quad     7.001\\

\hline
\end{tabular}
\end{center}
\end{table}

\begin{table}
\begin{center}
\begin{tabular}{p{7cm}}

\footnotesize Table 7. \footnotesize Quadrupole moments for $
_{45}^{103}Rh$ and $ _{56}^{135}Ba$ isotopes. The experimental
data for $ _{45}^{103}Rh$
\footnotesize isotope are taken from \cite{21,25}.The experimental data for $ _{56}^{135}Ba$   isotope are taken from \cite{21,31}.\\
\end{tabular}

\begin{tabular}{ccc}

\hline
Nucleus      \quad\quad\quad &$ J^{\pi}  $    \quad\quad & $ \frac {Q(eb)}{exp. \quad\quad\quad calc.} $    \\
\hline
$ _{45}^{103}Rh$   \quad\quad&  \,\,\, $(1/2)_{1}^{-} $        \quad\quad&    0                       \quad\quad \quad    0  \\
 \quad\quad &  $(3/2)_{1}^{-}  $  & -0.32                 \quad     -0.2588\\
\quad\quad &   $(5/2)_{1}^{-}  $  &-0. 41                       \quad     - 0.2824 \\
\quad\quad &   $(7/2)_{1}^{-} $  &   -             \quad\quad     -0.5538 \\
\quad\quad &   $(9/2)_{1}^{-} $  &-                \quad\quad    -0.5355 \\

$ _{56}^{135}Ba$   &   $ (3/2)_{1}^{+}$        &   0.146                               \quad   0.1509   \\
 \quad\quad &  $(1/2)_{1}^{+}$  &-                 \quad\quad     0.1349\\
\quad\quad &  $(5/2)_{1}^{+}$  & -                \quad\quad      0.5126 \\

\hline
\end{tabular}
\end{center}
\end{table}
\subsection{two-neutron separation energies}
Shape phase transitions in nuclei can be studied experimentally
by considering the behavior of the ground state energies of a
series of isotopes, or, more conveniently, the behavior of the
two-neutron separation energies, $ S_{2n}$ \cite{2}.On the other
hand, the ground-state two-neutron separation energies, $
S_{2n}$, are observables very sensitive to the details of the
nuclear structure. The occurrence of continuities in the
behavior of two-neutron separation energies describe a
second-order shape-phase transition between spherical and $
\gamma- unstable $ rotor limits \cite{2,32}.In due to, we have
investigated the evolution of two-neutron separation energies
along the Ba and Rh isotopic chains by both experimental and
theoretical values, which have been presented in Fig.13.
 The binding energy as a function of proton and neutron number is given by \cite{2}
 \begin{equation}
E_{B} (N_{\pi},N_{\nu} )=E_{c}+A_{\pi} N_{\pi}+A_{\nu}
N_{\nu}+\frac {1}{2} B_{\pi} N_{\pi} (N_{\pi}-1)+\frac {1}{2}
B_{\nu} N_{\nu} (N_{\nu}-1)+CN_{\pi} N_{\nu}+E_{D}
(N_{\pi},N_{\nu} )
 \end{equation}
Where $N_{\pi}(N_{\nu})$ is the number of proton (neutron) bosons
in the valence shell, $E_{c}$ the contribution from the core and
$E_{D}$ is the contribution to the binding energy due to the
deformation. Using Eq. (35), one obtains the following relation
for the two-neutron separation energy\cite{2}:
  \begin{equation}
S_{2n} (N_{\pi},N_{\nu} )=E_{B} (N_{\pi},N_{\nu} )-E_{B}
(N_{\pi},N_{\nu}-1)=A_{n}+BN_{\pi}+C_{n} N_{\nu}+(E_{D}
(N_{\pi},N_{\nu} )-E_{D} (N_{\pi},N_{\nu}-1 ) )
 \end{equation}
Using the $S_{2n}$ empirical values for these isotopic chains
 \cite{21} we have extracted $A_{n}+B N_{\pi}=15.1857 ,20.526
 $ Mev and $C_{n}=1.9784 ,-1.7992    $ Mev for Ba and Rh,
respectively. Then, we obtained the two-neutron separation
energies, which are shown in Fig.13, together with the
experimental values. It can be seen from Fig.13 that exist
continuities (linear variation) in the behavior of two-neutron
separation energies thus the phase transition for Ba and Rh
isotopic chains is of second order .Our result Confirmed the
predictions of done in refs. \cite{2,32}, where they suggest a
linear variation of $S_{2n}$ with respect to the neutron number
for the $U(5)-SO(6)$ transitional region.
\begin{figure}[htb]
\begin{center}
\includegraphics[height=6cm]{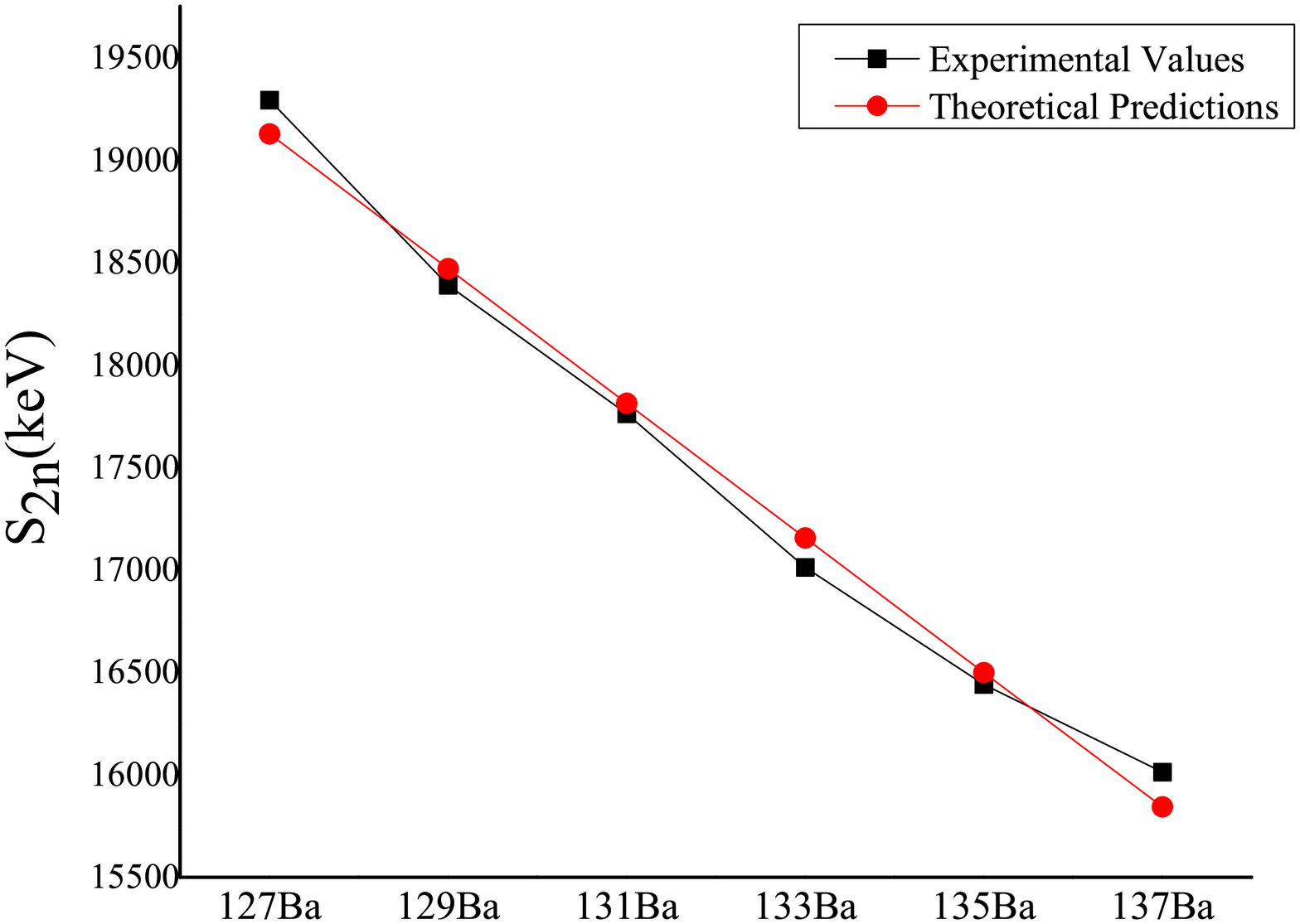}
\includegraphics[height=6cm]{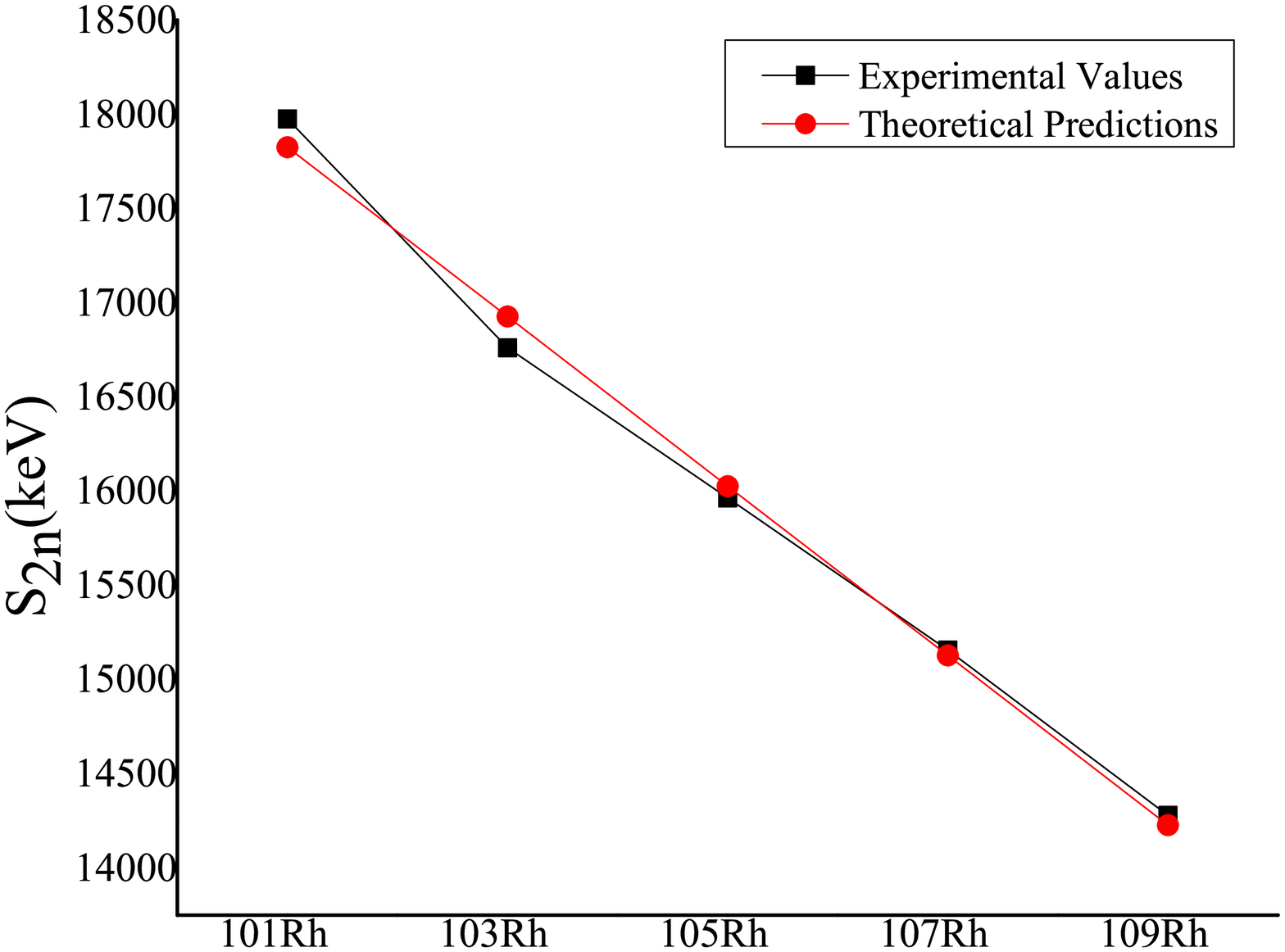}
\caption{A comparison between theoretical and experimental two
neutron separation energies,$S_{2n}$  (in keV) for Ba isotopes
(1eft panel) and Rh isotopes (right panel). Experimental data
from \cite{21}.\label{fig:11}}
\end{center}
\end{figure}

 \section{Conclusions}
In this paper, we have analyzed transition from spherical to
 $ \gamma -unstable $ shapes in odd -A nuclei. Key
observables of phase transition such as level crossing,
ground-state energy and derivative of the ground-state energy and
expectation values of the d-boson number operator have calculated.
We have presented experimental evidence for the $ U(5)-O(6)$
transition for negative parity states of the $ _{45}^{101-109}Rh$
isotopic chain and positive parity states of $ _{56}^{127-137}Ba$
isotopic chain, and performed an analysis for these isotopes via
a SU(1,1)-based Hamiltonian. The results indicate that the energy
spectra of the Rh and Ba isotopes can be reproduced quite well.
The calculated B(E2)values and two neutron separation energies
are agreements with the available experimental data. Our results
show that Rh isotopes have gamma-unstable rotor features but the
vibrational character is dominant while a dominancy of dynamical
symmetry O(6) exist for Ba isotopic chain and also $
_{45}^{105}Rh$ and $ _{56}^{133-135}Ba$ isotopes are as the best
candidates for $U^{BF} (5)-O^{BF} (6)$ transition.

\end{document}